\documentclass[aps,preprint,superscriptaddress,nofootinbib,preprintnumbers,prd]{revtex4-1}

\usepackage{bm,amsmath,amsfonts,amssymb,slashed,graphicx,xspace,placeins,multirow}

\allowdisplaybreaks
\DeclareMathAlphabet{\mathpzc}{OT1}{pzc}{m}{it}

\usepackage{hyperref}
\usepackage{breakurl}
\usepackage{array}

\newcommand{\nn}{\nonumber}
\newcommand{\GeV}{\text{GeV}}

\def\d{{\rm d}}
\newcommand{\ov}{\overline}

\newcommand{\g}{\gamma}
\newcommand{\dds}{D^{(*)}}

\newcommand{\cbar}{\bar{c}}
\newcommand{\Bbar}{\,\overline{\!B}{}}

\newcommand{\lqcd}{\ensuremath{\Lambda_{\rm QCD}}\xspace}
\newcommand{\aS}{\alpha_s}
\newcommand{\haS}{{\hat{\alpha}_s}}
\newcommand{\wm}{w_{z}}

\newcommand{\bddstn}{{\Bbar \to D^{(*)}\tau \bar\nu}}
\newcommand{\bdln}{{\Bbar \to D l \bar\nu}}
\newcommand{\bdelln}{{\Bbar \to D \ell \bar\nu}}
\newcommand{\bdsln}{{\Bbar \to D^{*} l \bar\nu}}
\newcommand{\bddsln}{{\Bbar \to D^{(*)} l \bar\nu}}
\newcommand{\bddselln}{{\Bbar \to D^{(*)} \ell \bar\nu}}

\newcommand{\ampBb}[2]{\big\langle #1 \big|\, #2\, \big| \Bbar \big\rangle}
\renewcommand{\ampBb}[2]{\langle #1 |\, #2\, | \Bbar \rangle}
\newcommand{\rD}{r_{D}}
\newcommand{\rDs}{r_{D^*}}
\newcommand{\brhosq}{\bar\rho_*^2}

\newcommand{\lqcdGF}{\text{L}_{w = 1}}
\newcommand{\lqcdGFQ}{\lqcdGF{+\text{SR}}}
\newcommand{\normDDs}{\text{NoL}}
\newcommand{\normDDsQ}{\normDDs{+\text{SR}}}
\newcommand{\lqcdXw}{\text{L}_{w \ge 1}} 
\newcommand{\lqcdXwQ}{\lqcdXw{+\text{SR}}}
\newcommand{\lqcdXwQth}{\text{th:}\lqcdXw{+\text{SR}}}

\makeatletter
\g@addto@macro\bfseries{\boldmath}
\makeatother

\begin{document}

\title{Combined analysis of semileptonic $B$ decays to $D$ and $D^*$:
$R(D^{(*)})$, $|V_{cb}|$, and new physics}

\author{Florian U.\ Bernlochner}
\affiliation{Physikalisches Institut der Rheinischen Friedrich-Wilhelms-Universit\"at Bonn, 53115 Bonn, Germany}

\author{Zoltan Ligeti}
\affiliation{Ernest Orlando Lawrence Berkeley National Laboratory, 
University of California, Berkeley, CA 94720, USA}

\author{Michele Papucci}
\affiliation{Ernest Orlando Lawrence Berkeley National Laboratory, 
University of California, Berkeley, CA 94720, USA}

\author{Dean J.\ Robinson}
\affiliation{Physics Department, University of Cincinnati, Cincinnati OH 45221, USA}

\begin{abstract}
The measured $\bar{B}\to D^{(*)} l \bar\nu$ decay rates for light leptons ($l=e,\mu$)
constrain all $\bar{B}\to D^{(*)}$ semileptonic form factors, by including both the
leading and ${\cal O}(\Lambda_{\text{QCD}}/m_{c,b})$ subleading Isgur-Wise functions in the
heavy quark effective theory.  We perform a novel combined fit to the $\bar{B}\to
D^{(*)} l \bar\nu$ decay distributions to predict the $\bar{B} \to D^{(*)} \tau\bar\nu$
rates and determine the CKM matrix element $|V_{cb}|$.  Most theoretical and
experimental papers have neglected uncertainties in the predictions for form
factor ratios at order $\Lambda_{\text{QCD}}/m_{c,b}$, which we include.  We also calculate
${\cal O}(\Lambda_{\text{QCD}}/m_{c,b})$ and ${\cal O}(\alpha_s)$ contributions to semileptonic
$\bar{B}\to D^{(*)} \ell \bar\nu$ decays for all possible $b \to c$ currents.  This result has not
been available for the tensor current form factors, and for two of those, which
are ${\cal O}(\Lambda_{\text{QCD}}/m_{c,b})$, the corrections are of the same order as approximations used in the
literature.  These results allow us to determine with improved
precision how new physics may affect the $\bar{B}\to D^{(*)} \tau\bar\nu$ rates.  Our
predictions can be systematically improved with more data; they need not rely on
lattice QCD results, although these can be incorporated.
\end{abstract}

\maketitle

\section{introduction}

Heavy quark symmetry~\cite{Isgur:1989vq, Isgur:1989ed} plays an essential role
in understanding exclusive semileptonic $b\to c\ell\bar\nu$ mediated
transitions, by providing relations between hadronic form factors.  At leading
order in $\lqcd/m_{c,b}$, the symmetry also determines the absolute
normalization of form factors at the ``zero recoil" point, $v_B = v_{D^{(*)}}$,
corresponding to maximal invariant mass, $q^2$,  of the outgoing lepton pair.
Incorporating small corrections to the symmetry limit permits a (hadronic)
model-independent determination of $|V_{cb}|$ from exclusive decays.  Recently,
the Babar~\cite{Lees:2012xj, Lees:2013uzd}, Belle~\cite{Huschle:2015rga,
Abdesselam:2016cgx, Abdesselam:2016xqt}, and LHCb~\cite{Aaij:2015yra}
measurements of the $|V_{cb}|$-independent ratios
\begin{equation}
  R(\dds) = \frac{\Gamma(\bddstn)}{\Gamma(\bddsln)}\,, \qquad l = \mu, \,e\,,
\end{equation}
renewed interest in these decays.  The world average of $R(D)$ and $R(D^*)$ is
in tension with the SM expectation at the $4\sigma$ level~\cite{HFAG}.  This is
intriguing as it occurs in a tree-level SM process, while most new physics
(NP) explanations require new states at or below one
TeV~\cite{Freytsis:2015qca}. 

Besides the search for new physics, understanding $b\to c\ell\bar\nu$
mediated semileptonic decays as precisely as possible is also important for
future improvements of the determinations of the CKM elements $|V_{cb}|$ and
$|V_{ub}|$, both from exclusive and inclusive $B$ decays, which exhibit some
tensions~\cite{HFAG}.  Depending on the particular measurement, some decay modes
contribute to the signals, some to the backgrounds.  Future progress is
essential for increasing the scale of new physics probed by the Belle~II and
LHCb experiments~\cite{Charles:2013aka}. 

The main uncertainty in predicting $R(\dds)$ comes from the fact that the
$\bddstn$ decay rates depend on certain form factors, that only give $m_l^2/m_B^2$
suppressed contributions to the differential rates for the precisely measured
light lepton channels. Using heavy quark effective theory (HQET), however, all
$\Bbar \to \dds$ form factors are described by a single Isgur-Wise function in
the $m_{c,b}\gg \lqcd$ limit.  At order $\lqcd / m_{c,b}$, only three additional
functions of $q^2$ are needed to parametrize all form factors. 

We perform the first combined fit to $\bddsln$ differential rates
and angular distributions, including  $\mathcal{O}(\lqcd/{m_{c,b}},\, \aS)$
terms in HQET, to constrain both the leading and three subleading Isgur-Wise
functions. This fit constrains all form factors, up to higher order corrections,
with uncertainties suppressed by $\mathcal{O}(\lqcd^2/m^2_{c,b}\,,\,\aS\lqcd
/m_{c,b}\,,\, \aS^2)$.  We extract $|V_{cb}|$ and form factor ratios under
various fit scenarios, that include or omit lattice QCD and/or QCD sum rule
inputs, and which provide checks of previously  untested theory assumptions or
results. Most prior theoretical and experimental studies neglected HQET relations for the
form factors at order $\lqcd/m_{c,b}$ or the correlations of the uncertainties
in the deviations from the heavy quark limit.  
Our fits fully incorporate these.  These fits also allow precise predictions
of the $\bddstn$ rates and $R(D^{(*)})$. Our predictions can be systematically
improved with more $\bddsln$ data, and need not rely on lattice  QCD results. A
similar approach to analyze $\Bbar \to D^{**} l\bar\nu$ decays was recently
carried out in Ref.~\cite{Bernlochner:2016bci}.

We also compute, for all possible $b \to c$ currents, the
$\mathcal{O}(\lqcd/m_{c,b})$ and $\mathcal{O}(\aS)$ contributions to the form
factors. While the $\mathcal{O}(\lqcd/m_{c,b})$ corrections to the vector and
axial-vector matrix elements have been known for over 25
years~\cite{Luke:1990eg, Neubert:1991xw}, the corrections for the tensor current
form factors are not explicitly available in past literature.  Two of these form
factors vanish in the heavy quark limit, and receive unsuppressed  corrections
to the partial results, also of order $\lqcd/m_{c,b}$, used previously in the
literature.

Section~\ref{sec:hqet} contains the HQET calculations of the form factors,
including order $\lqcd/m_{c,b}$ and $\aS$ contributions, corresponding
expressions for form factor ratios, and some details of our numerical
evaluations in the $1S$ scheme to avoid known bad behaviors in the perturbation
expansions.  In Section \ref{sec:fit} we review analyticity constraints on the
form factors, parametrizations of the Isgur-Wise functions, and develop several fit scenarios consistent with HQET, which we
apply to the data. The results for $|V_{cb}|$, form factor ratios,
and $R(\dds)$ are discussed. Section~\ref{sec:conc} concludes.

\section{Elements of HQET}
\label{sec:hqet}

\subsection{Matrix elements to order \texorpdfstring{$\lqcd/m_{c,b}$}{lqcdm} and \texorpdfstring{$\aS$}{as}}
\label{sec:form}

We are concerned with matrix elements $\langle D^{(*)} |\, O_\Gamma\,
| \Bbar \rangle$, where a full operator basis is
\begin{equation}\label{eqn:Gdef}
O_S = \bar c\, b\,, \qquad  O_P = \bar c\, \g^5\, b\,, \qquad
O_V = \bar c\,\g^\mu\, b\,, \qquad  O_A = \bar c\, \g^\mu\g^5\, b\,, \qquad
O_T = \bar c\, \sigma^{\mu\nu}\, b\,,
\end{equation}
with $\sigma^{\mu\nu} \equiv (i/2)[\g^\mu,\g^\nu]$.  (The sign convention is
fixed by $\sigma^{\mu\nu} \g^5 \equiv -(i/2)\epsilon^{\mu \nu \rho \sigma}
\sigma_{\rho \sigma}$, which implies
$\text{Tr}[\g^\mu\g^\nu\g^\sigma\g^\rho\g^5] = +4i
\epsilon^{\mu\nu\rho\sigma}$.)  The construction of the HQET expansion to order
$\mathcal{O}(\lqcd/m_{c,b})$ and $\mathcal{O}(\aS)$ was developed in the early
'90s~\cite{Manohar:2000dt, Neubert:1993mb}; we summarize here the central
elements to establish our conventions.

The HQET allows model independent parametrization of the spectroscopy of heavy
mesons and some hadronic matrix elements between them.  The ground state heavy
quark spin symmetry doublet pseudoscalar ($P$) and vector ($V$) mesons
correspond to the light degrees of freedom (the ``brown muck") in a
spin-$\frac12$ state combined with the heavy quark spin.  They form two states
with angular momentum $J_{V,P} = \frac12 \pm \frac12$.  Their masses can be
expressed as
\begin{equation}
m_{V,P} = m_Q + \bar\Lambda - \frac{\lambda_1}{2m_Q} \pm 
  \frac{(2J_{P,V}+1)\lambda_2}{2m_Q} + \ldots\,,
\end{equation}
where $m_Q$ is the heavy quark mass parameter of HQET, $\bar\Lambda = {\cal
O}(\lqcd)$, $\lambda_{1,2} = {\cal O}(\lqcd^2)$, etc.  To evaluate matrix
elements relevant for semileptonic decays, it is simplest to use the trace
formalism~\cite{Falk:1990yz, Bjorken:1990rr, Falk:1991nq}. Including
$\lqcd/m_{c,b}$ corrections, the $B\to D^{(*)}$ matrix elements can be written
as~\cite{Falk:1992wt}
\begin{align}
\label{eqn:TraceLM}
\frac{\langle \dds |\, \cbar\, \Gamma\, b\, | \Bbar \rangle}{\sqrt{m_{\dds} m_B}}
  = - \xi(w)\, & \Big\{ \text{Tr} \big[ \bar H^{(c)}_{v'}\, \Gamma\,
  H^{(b)}_v \big] \nn \\
& + \varepsilon_c\, \text{Tr}\big[ \bar H^{(c,1)}_{v',v}\, 
  \Gamma\, H^{(b)}_v \big] 
  + \varepsilon_b\, \text{Tr} \big[ \bar H^{(c)}_{v'}\, \Gamma\, 
  H^{(b,1)}_{v,v'} \big]\Big\}\,, 
\end{align}
where $\varepsilon_{c,b} = \bar\Lambda/(2m_{c,b})$ and $\Gamma$ is an
arbitrary Dirac matrix.  The pseudoscalar and vector mesons can be represented
by a ``superfield", which has the right transformation properties under heavy
quark and Lorentz symmetry,
\begin{equation}\label{eqn:Hexp0}
H_v^{(Q)} = \frac{1+\slashed v}{2}\, \big( V_v^{(Q)} \slashed{\epsilon}
   - P_v^{(Q)} \g_5 \big) \,.
\end{equation}
The $\lqcd/m_{c,b}$ corrections can be parametrized via~\cite{Falk:1992wt}
\begin{align}\label{eqn:Hexp1}
H_{v,v'}^{(Q,1)} & = \frac{1+\slashed v}{2}\,
  \Big\{ V_v^{(Q)} \big[\slashed{\epsilon} \hat{L}_2(w) 
  + \epsilon \cdot v' \hat{L}_3(w) \big] - P_v^{(Q)}\, \g_5\, \hat{L}_1(w) \Big\} \nn \\
& + \frac{1-\slashed v}{2}\, \Big\{ V_v^{(Q)} \big[ \slashed\epsilon \hat{L}_5(w) 
  + \epsilon \cdot v' \hat{L}_6(w) \big] - P_v^{(Q)}\, \g_5\, \hat{L}_4(w) \Big\}\,.
\end{align}
It is convenient to use the dimensionless kinematic variable $w$ instead of $q^2 = (p_B -
p_{\dds})^2$,
\begin{equation}\label{wdef}
w = v\cdot v' = \frac{m_B^2  + m_{\dds}^2 - q^2}{2m_B m_{\dds}}\,,\qquad
  v = \frac{p_B}{m_B}\,, \qquad v' = \frac{p_{\dds}}{m_{\dds}}\,.
\end{equation}
In Eq.~\eqref{eqn:TraceLM} and hereafter, we absorb into the leading order
Isgur-Wise function a heavy quark spin symmetry conserving ${\cal
O}(\lqcd/m_{c,b})$ subleading term, which does not affect any model independent
predictions of HQET, via $\xi(w) \to \xi(w) + 2 (\varepsilon_c + \varepsilon_b)
\chi_1(w)$.  The function $\chi_1$ parametrizes the matrix element of the time
ordered product of the kinetic operator in the subleading HQET Lagrangian,
$O_{\rm kin} = \bar h_v\,(i D)^2\, h_v /(2m_Q)$, with the leading order
current.  It satisfies $\chi_1(1)=0$~\cite{Luke:1990eg}, and hence $\xi(1) = 1$
is maintained.  Reparametrization invariance~\cite{Luke:1992cs} ensures that
this redefinition of $\xi(w)$ is RGE invariant.

The $w$-dependent $L_{1\ldots6}$ functions are~\cite{Falk:1992wt}
\begin{align}\label{Lhatdef}
\hat{L}_1 &= - 4(w-1) \hat\chi_2 + 12 \hat\chi_3\,, \qquad 
  \hat{L}_2 = - 4 \hat\chi_3\,, \qquad \hat{L}_3 = 4 \hat\chi_2\,, \nn\\*
\hat{L}_4 &= 2 \eta - 1 \,, \qquad \hat{L}_5 = -1\,, \qquad \hat{L}_6 = - 2 (1 + \eta)/(w+1)\,.
\end{align}
Here the $\hat{\chi}_{2,3}$ terms in $\hat{L}_{1,2,3}$ originate from the matrix
elements of the time ordered product of the leading order current with the
chromomagnetic correction to the Lagrangian, $O_{\rm mag} = (g_s/2)\, \bar
h_v\sigma_{\mu\nu}G^{\mu \nu} h_v/(2m_Q)$.  Luke's theorem implies
$\hat{\chi}_3(1) = 0$~\cite{Luke:1990eg}.  The $\hat{L}_{4,5,6}$ terms arise
from $\lqcd/m_{c,b}$ corrections in the matching of the $\cbar\, \Gamma b$ heavy
quark current onto HQET, $\cbar\, \Gamma b \to \cbar_{v'} \big[ \Gamma -
i\overleftarrow{\slashed{D}}\, \Gamma/(2m_c) + \Gamma\,
i\overrightarrow{\slashed{D}}/(2m_b) + \ldots\big]b_v$.\footnote{Our definitions
of the subleading Isgur-Wise functions, $\chi_{1,2,3}$, $\eta$, and hence
$\hat{L}_{1\ldots 6}$, are dimensionless due to factoring out $\bar\Lambda$, as
done, e.g., in Refs.~\cite{Neubert:1993mb, Grinstein:2001yg} but not in
Refs.~\cite{Luke:1990eg, Manohar:2000dt}; the correspondence is obvious.  The
QCD sum rule calculations~\cite{Neubert:1992wq, Neubert:1992pn, Ligeti:1993hw}
also compute these functions with the dimensionless definitions.}

The perturbative corrections to the heavy quark currents may be computed by
matching QCD onto HQET~\cite{Falk:1990yz, Falk:1990cz, Neubert:1992qq}.  At
$\mathcal{O}(\aS)$, the following operators are generated
\begin{align}
\label{eqn:ascurrent}
\cbar\, b		&\to  \cbar_{v'} \big( 1+ \haS\, C_S\big) b_v\,, \nn\\
\cbar \gamma^5 b	&\to  \cbar_{v'} \big(1+ \haS\, C_P \big) \g^5 b_v\,,\nn\\
\cbar \g^\mu b		&\to  \cbar_{v'} \big[\big(1+ \haS\, C_{V_1} \big) \g^\mu + \haS\, C_{V_2}\, v^\mu + \haS\, C_{V_3}\, v'^\mu \big]b_v \,, \nn\\
\cbar \g^\mu\g^5 b	&\to  \cbar_{v'} \big[\big(1+ \haS\, C_{A_1} \big) \g^\mu + \haS\, C_{A_2}\, v^\mu + \haS\, C_{A_3}\, v'^\mu \big] \g^5 b_v\, \nn \\
\cbar \sigma^{\mu\nu} b	&\to \cbar_{v'} \big[\big(1+ \haS\, C_{T_1} \big) \sigma^{\mu\nu} + \haS\, C_{T_2}\, i(v^\mu\g^\nu - v^\nu\g^\mu) 
  + \haS\, C_{T_3}\, i(v'^\mu\g^\nu - v'^\nu\g^\mu)  \nn \\
  & \qquad\quad + C_{T_4}(v'^\mu v^\nu - v'^\nu v ^\mu)\big] b_v\,,
\end{align}
where the $C_{\Gamma_i}$ are functions of $w$ and $z=m_c/m_b$, and $\haS = \aS/\pi$. (We
follow the notation of Ref.~\cite{Manohar:2000dt}, while
Ref.~\cite{Neubert:1993mb} uses $C_i = \haS C_{V_i} + \delta_{i1}$ and $C_i^5 =
\haS C_{A_i} + \delta_{i1}$.)  Evaluating these contributions using the leading
order trace in Eq.~\eqref{eqn:TraceLM} leads to  $\mathcal{O}(\aS)$
modifications of the coefficients of the Isgur-Wise function, $\xi(w)$.  In this
paper we neglect $\mathcal{O}(\aS\, \varepsilon_{c,b})$ corrections, which can
also be included straightforwardly (and should be, if NP is established).

The $\aS$ corrections for all five currents were computed in
Ref.~\cite{Neubert:1992qq}.  Appendix~\ref{app:aSC} contains their explicit
expressions, at arbitrary matching scale $\mu$. The vector and axial-vector
currents are not renormalized in QCD, but the corresponding heavy quark currents
have non-zero anomalous dimensions, leading to $\mu$-dependence for $C_{V_1}$
and $C_{A_1}$ for $w \neq 1$. The scalar, pseudoscalar, and tensor currents are
renormalized in QCD, and thus $C_S$, $C_P$, and $C_{T_1}$ are also
$\mu$-dependent. In the $\overline{\rm MS}$ scheme with dimensional
regularization, the remaining $C_{\Gamma_j}$ ($j\ge 2$) are scale independent.

\subsection{\texorpdfstring{$\Bbar \to \dds$}{BDDs} form factors}
\label{ffsubsec}

We use the standard definitions of the form factors. For $\Bbar \to D$ decays, 
\begin{subequations}
\begin{align}
\ampBb{D}{\cbar\,b} & = \sqrt{m_B m_D}\, h_S\, (w+1)\,, \label{eqn:DS} \\*
\ampBb{D}{\cbar\g^5 b} & = \ampBb{D}{\cbar \g^\mu\g^5 b} = 0\,, \label{eqn:DPA} \\
\ampBb{D}{\cbar \g^\mu b} & = \sqrt{m_B m_D}\, 
  \big[ h_+(v+v')^\mu + h_-(v-v')^\mu\big], \\*
\ampBb{D}{\cbar \sigma^{\mu\nu} b} & = i \sqrt{m_B m_D}\, 
  \big[ h_T\, (v'^\mu v^\nu - v'^\nu v^\mu )\big],
\end{align}
\end{subequations}
while for the $\Bbar \to D^*$ transitions, 
\begin{subequations}
\begin{align}
\ampBb{D^*}{\cbar b} & = 0\,, \label{eqn:DsS}\\*
\ampBb{D^*}{\cbar \g^5 b} & = -\sqrt{m_B m_{D^*}}\, h_P\, (\epsilon^* \cdot v)\,, \\
\ampBb{D^*}{\cbar \g^\mu b} & = i\sqrt{m_B m_{D^*}}\, h_V\, \varepsilon^{\mu\nu\alpha\beta}\,
  \epsilon^*_{\nu}v'_\alpha v_\beta \,,\label{eqn:DsV}\\
\ampBb{D^*}{\cbar \g^\mu \g^5 b} & = \sqrt{m_B m_{D^*}}\, \big[h_{A_1} (w+1)\epsilon^{*\mu}
  - h_{A_2}(\epsilon^* \cdot v)v^\mu 
  - h_{A_3}(\epsilon^* \cdot v)v'^\mu \big] ,\label{eqn:DsA}\\
\ampBb{D^*}{\cbar \sigma^{\mu\nu} b} & = -\sqrt{m_B m_{D^*}}\,
  \varepsilon^{\mu\nu\alpha\beta} \big[ h_{T_1} \epsilon^*_{\alpha}
  (v+v')_\beta + h_{T_2} \epsilon^*_{\alpha} (v-v')_\beta 
  + h_{T_3}(\epsilon^*\cdot v) v_\alpha v'_\beta\big]. \label{eqn:DsT}
\end{align}
\end{subequations}
The $i$, $-1$, and $w+1$ factors are chosen such that in the heavy quark
limit each form factor either vanishes or equals the leading order Isgur-Wise
function,
\begin{gather}\label{BDhlead}
	h_-  =  h_{A_2} = h_{T_2} = h_{T_3} = 0\,, \nn\\
	h_+  = h_V = h_{A_1} = h_{A_3} = h_S = h_P = h_T = h_{T_1} = \xi\,.
\end{gather}

Using Eqs.~\eqref{eqn:TraceLM} and \eqref{eqn:ascurrent}, one can compute
all form factors to order $\mathcal{O}(\lqcd/m_{c,b})$ and $\mathcal{O}(\aS)$.
It is convenient to factor out $\xi(w)$, defining
\begin{equation}\label{eqn:hatHdef}
	\hat{h}(w) = h(w) / \xi(w) \,.
\end{equation}
By virtue of Eq.~\eqref{eqn:Hexp1}, the $\bdln$ form factors only depend on two
linear combinations of subleading Isgur-Wise functions, $\hat{L}_1$ and
$\hat{L}_4$,
\begin{align}\label{eqn:BD1m}
\hat h_+ & = 1 + \haS\Big[C_{V_1} + \frac{w+1}2\, (C_{V_2}+C_{V_3})\Big]
  + (\varepsilon_c + \varepsilon_b)\, \hat{L}_1 \,, \nn\\*
\hat h_- & = \haS\, \frac{w+1}2\, (C_{V_2}-C_{V_3}) + (\varepsilon_c - \varepsilon_b)\, \hat{L}_4 \,, \nn\\
\hat h_S & = 1 + \haS\, C_S + (\varepsilon_c + \varepsilon_b) 
  \bigg(\! \hat{L}_1 - \hat{L}_4\, \frac{w-1}{w+1} \bigg)\,, \nn\\
\hat h_T & = 1 + \haS \big(C_{T_1}-C_{T_2}+C_{T_3}\big) + (\varepsilon_c + \varepsilon_b) \big( \hat{L}_1 - \hat{L}_4 \big) \,.
\end{align}
For the $\bdsln$ form factors we obtain
\begin{align}\label{eqn:BDs1m}
\hat h_V 	& = 1 + \haS\, C_{V_1} + \varepsilon_c \big(\hat{L}_2 - \hat{L}_5\big)  + \varepsilon_b \big( \hat{L}_1 - \hat{L}_4 \big) \,,\nn\\
\hat h_{A_1} 	& = 1 + \haS\, C_{A_1} 
  + \varepsilon_c \bigg(\! \hat{L}_2 - \hat{L}_5\, \frac{w-1}{w+1} \bigg)
  + \varepsilon_b \bigg(\! \hat{L}_1 - \hat{L}_4\, \frac{w-1}{w+1} \bigg) \,,\nn\\
\hat h_{A_2} 	& = \haS\, C_{A_2} + \varepsilon_c \big(\hat{L}_3 + \hat{L}_6\big) \,,\nn\\
\hat h_{A_3} 	& = 1 + \haS \big(C_{A_1} + C_{A_3}\big) + \varepsilon_c \big(\hat{L}_2 - \hat{L}_3 + \hat{L}_6 - \hat{L}_5 \big) + \varepsilon_b \big(\hat{L}_1 - \hat{L}_4\big) \,,\nn\\
\hat h_P 	& = 1 + \haS\, C_P + \varepsilon_c \big[\hat{L}_2 + \hat{L}_3 (w-1)  + \hat{L} _5 - \hat{L}_6(w+1)\big]  + \varepsilon_b \big( \hat{L}_1 - \hat{L}_4 \big) \,,\nn\\
\hat h_{T_1} 	& = 1 + \haS \Big[ C_{T_1}  + \frac{w-1}2\, \big(C_{T_2}-C_{T_3}\big) \Big] + \varepsilon_c \hat{L}_2 + \varepsilon_b \hat{L}_1 \,,\nn\\
\hat h_{T_2} 	& = \haS\, \frac{w+1}2\, \big(C_{T_2}+C_{T_3}\big) + \varepsilon_c \hat{L}_5 - \varepsilon_b \hat{L}_4 \,,\nn\\
\hat h_{T_3} 	& = \haS\, C_{T_2} + \varepsilon_c \big(\hat{L}_6 - \hat{L}_3\big) \,.
\end{align}
In Eqs.~\eqref{eqn:BD1m} and \eqref{eqn:BDs1m}, the relations for the SM
currents --- that is, $h_+$, $h_-$, $h_V$, $h_{A_1}$, $h_{A_2}$, and $h_{A_3}$
--- agree with the literature, e.g., Refs.~\cite{Falk:1992wt, Neubert:1993mb}.
Because of Luke's theorem, the $\mathcal{O}(\lqcd/m_{c,b})$ corrections to
$h_+$, $h_S$, $h_{A_1}$, and $h_{T_1}$ vanish at zero recoil.  To the best of
our knowledge, the expressions for $h_T$ and $h_{T_{1,2,3}}$ cannot be
found in the literature.  For $h_{T_2}$ and $h_{T_3}$, which start at order
$\lqcd/m_{c,b}$, the partial results used in the literature (e.g.,
Ref.~\cite{Tanaka:2012nw}) kept and left out terms, which are both order
$\mathcal{O}(\lqcd/m_{c,b})$.

The scalar and vector matrix elements in $\Bbar\to D$ transitions, and the
pseudoscalar and axial vector ones in $\Bbar\to D^*$, are related by the
equations of motion
\begin{gather}
	[\ov m_b(\mu)-\ov m_c(\mu)]\,\ampBb{D}{\cbar\,b} =  \ampBb{D}{ \cbar\, \slashed{q}\, b }\,, \nn\\
	-[\ov m_b(\mu)+\ov m_c(\mu)]\,\ampBb{D^*}{\cbar \g^5 b} = \ampBb{D^*}{ \cbar \, \slashed{q} \g^5\,  b }\,,\label{currentrel}
\end{gather}
in which $\ov m_Q(\mu)$ are the $\overline{\rm MS}$ quark masses at a common
scale $\mu$, obeying
\begin{equation}
  m_Q = \ov m_Q(\mu) \bigg[ 1+ \haS \bigg( \frac43 - \ln \frac{m_Q^2}{\mu^2}
  \bigg) + \ldots \bigg] \,.
\end{equation}
One can verify using $m_b = m_B - \bar\Lambda + \mathcal{O}(\lqcd^2/m_b)$ and
$m_c = m_{D^{(*)}} - \bar\Lambda + \mathcal{O}(\lqcd^2/m_c)$ that the form
factor expansions in Eqs.~\eqref{eqn:BD1m} and \eqref{eqn:BDs1m} satisfy these
relations, including all $\mathcal{O}(\varepsilon_{c,b})$ and $\mathcal{O}(\aS)$
terms.  We emphasize that this only holds using the $\overline{\rm MS}$ masses
at the common scale $\mu$.  Using $\ov m_b(\ov m_b)$ and $\ov m_c(\ov
m_c)$~\cite{PDG} in Eqs.~\eqref{currentrel}, as done in some papers, is
inconsistent.

We prefer to evaluate the scalar and pseudoscalar matrix elements using
Eqs.~(\ref{eqn:BD1m}) and (\ref{eqn:BDs1m}) instead of Eq.~(\ref{currentrel}),
because the natural choice for $\mu$ is below $m_b$ (or sometimes well below, as
in the small-velocity limit~\cite{Shifman:1987rj, Boyd:1995ht}).  
In the $\ov{\rm MS}$ scheme fermions do not decouple for $\mu < m$, introducing
artificially large corrections in the running, compensated by corresponding
spurious terms in the $\beta$-function computed without integrating out heavy
quarks~\cite{Manohar:1996cq}.

\subsection{Decay rates and form factor ratios}

The $\bddsln$ differential rates have the well-known expressions in the SM,
\begin{subequations}
\begin{align}
\frac{\d \Gamma(\bdln)}{\d w} & = \frac{G_F^2|V_{cb}|^2 \, \eta_{\rm EW}^{2} \, m_B^5}{48 \pi^3}\,
  (w^2-1)^{3/2}\, \rD^3\, (1 + \rD)^2\, \mathcal{G}(w)^2\,,\\
\frac{\d \Gamma(\bdsln)}{\d w} & = \frac{G_F^2|V_{cb}|^2 \, \eta_{\rm EW}^{2} \,  m_B^5}{48 \pi^3}\,
   (w^2-1)^{1/2}\, (w + 1)^2\, \rDs^3 (1- \rDs)^2 \nn \\
  & \quad \times \bigg[1 + \frac{4w}{w+1}\frac{1- 2 w\rDs + \rDs^2}{(1 - \rDs)^2} \bigg] \mathcal{F}(w)^2\,,
\end{align}
\end{subequations}
where $r_{\dds} = m_{\dds}/m_B$ and $\eta_{\rm EW} \simeq
1.0066$~\cite{Sirlin:1981ie}  is the electroweak correction.  In addition,
\begin{subequations}
\label{eqn:GFdef}
\begin{align}
\mathcal{G}(w) & = h_+ - \frac{1-\rD}{1 + \rD} h_- \,,\\
\mathcal{F}(w)^2 & = h_{A_1}^2 \bigg\{ 2(1 - 2 w \rDs + \rDs^2)
  \bigg(1 + R_1^2\, \frac{w-1}{w+1}\bigg) 
  + \big[(1 - \rDs) + (w-1)\big( 1 - R_2\big) \big]^2 \bigg\} \nn \\
	& \qquad\quad \times \bigg[(1 - \rDs)^2 + \frac{4w}{w+1}\big(1- 2 w\rDs + \rDs^2\big) \bigg]^{-1}\,,
\end{align}
\end{subequations}
and the form-factor ratios are defined as
\begin{equation}\label{eqn:R1R2Def}
	R_1(w) = \frac{h_V}{h_{A_1}}\,, \qquad 
	R_2(w) = \frac{h_{A_3} + \rDs\, h_{A_2}}{h_{A_1}}\,.
\end{equation}
In the heavy quark limit, $R_{1,2}(w) = 1$ and $\mathcal{F}(w) =
\mathcal{G}(w) = \xi(w)$, the leading Isgur-Wise function. It is common to fit
the measured $\bdsln$ angular distributions to $R_{1,2}(w)$. To
$\mathcal{O}(\varepsilon_{c,b},\, \aS)$, the SM predictions are
\begin{align}
	\label{eqn:R12exp}
R_1(w) & = 1 + \haS\big(C_{V_1} - C_{A_1}\big) - \frac{2}{w+1}\,
  \big(\varepsilon_b \hat{L}_4 + \varepsilon_c \hat{L}_5 \big)\,, \\*
R_2(w) & = 1 + \haS\big(C_{A_3} + \rDs C_{A_2}\big) - 
  \frac{2}{w+1}\, \big(\varepsilon_b \hat{L}_4 + \varepsilon_c \hat{L}_5 \big) 
  + \varepsilon_c\big[ \hat{L}_6(1 + \rDs) - \hat{L}_3(1 - \rDs) \big]\,. \nn 
\end{align}

To include the lepton mass suppressed terms, one sometimes
defines~\cite{Fajfer:2012vx,Tanaka:2012nw} additional form factor ratios
\begin{equation}
  R_3(w) = \frac{h_{A_3} - \rDs h_{A_2}}{h_{A_1}}\,, \qquad
  R_0(w) = \frac{h_{A_1}(w+1) - h_{A_3}(w -\rDs) - h_{A_2}(1 - w\rDs)}
  {(1+r_{D^*})\, h_{A_1}}\,.
\end{equation}
All contributions of $R_{0,3}(w)$ are proportional to $m_\ell^2$. (Ref.~\cite{Fajfer:2012vx} defines $R_3 = h_{A_3}/h_{A_1}$.) They are not linearly independent from $R_{1,2}(w)$, as there are only three form factor ratios
in $B\to D^*\ell\bar\nu$ in the SM.  In the heavy quark limit, $R_3(w) = R_0(w) =1$.  At
$\mathcal{O}(\varepsilon_{c,b}, \, \aS)$, the SM predictions are
\begin{align}
	R_3(w) & = 1 + \haS\big(C_{A_3} - \rDs C_{A_2}\big) - \frac{2}{w+1}\, \big(\varepsilon_b \hat{L}_4 + 
			\varepsilon_c \hat{L}_5 \big)  + \varepsilon_c\big[ \hat{L}_6(1 - \rDs) - \hat{L}_3(1 + \rDs) \big]\,, \nn \\
  	R_0(w) & = 1 + \haS\frac{C_{A_3}(\rDs-w) - (1-\rDs w) C_{A_2}}{1 + \rDs} + \frac{2(w -\rDs)}{(1 + \rDs)(1+ w)}\big( \varepsilon_b \hat{L}_4 + \varepsilon_c \hat{L}_5\big)  \nn \\
			& \qquad + \varepsilon_c\bigg[\hat{L}_3(w-1) - \hat{L}_6(w+1)\frac{1-\rDs}{1+ \rDs} \bigg]\,.
\end{align}

\subsection{The \texorpdfstring{$1S$}{1S} scheme and numerical results}
\label{sec:NV}

The $C_{\Gamma}$ coefficients defined in Eq.~\eqref{eqn:ascurrent} are
functions of $w$ and $z = m_c/m_b$, and thus depend on the quark masses.   As
is well known, the pole mass of a heavy quark contains a leading renormalon
ambiguity of order $\lqcd$, and so does the HQET parameter $\bar\Lambda$, as
they are ill-defined beyond perturbation theory.  The ambiguity is canceled by
a corresponding ambiguity in the perturbation series, connected to factorial
growth of the coefficients of $\haS^n$~\cite{Neubert:1994wq, Luke:1994xd, Beneke:1994sw,Bigi:1994em,Beneke:1994bc}.  The
cancellation comes about as a non-analytic term connected to the asymptotic
nature of the perturbation series, $e^{-c/\aS(M)} \sim
(\lqcd/M)^{c\beta_0/(4\pi)}$, where $\beta_0 = (11- 2 n_f/3)$ is the first
coefficient in the expansion of the $\beta$ function.  For example,
Eq.~\eqref{eqn:R12exp} implies at zero recoil, $R_1(1) \simeq 1 + 4\haS/3 +
\varepsilon_c + \varepsilon_b - 2\varepsilon_b\eta(1)$, where the order
$\haS^2\beta_0$ terms are also known~\cite{Grinstein:2001yg}.  The leading
renormalon corresponding to the worst behavior of the $\haS^n$ power series is
canceled by the ambiguity in $\bar\Lambda$ within the $\varepsilon_c +
\varepsilon_b$ term. The $-2\varepsilon_b\eta(1)$ term, however, does not
contribute to this leading renormalon cancellation, as the only participating terms are those
$\bar\Lambda/m_{c,b}$ terms not multiplied by any subleading Isgur-Wise
functions.

The $\aS$ perturbation series is known to be poorly convergent for many $B$
decay processes already at $\mathcal{O}(\aS^2)$, when expressed in terms of the
pole mass. To ensure the order-by-order cancellation of the fastest factorially
growing terms, it is convenient to reorganize the perturbation series in terms
of a suitable short-distance mass scheme, instead of the pole mass.  We use the
$1S$  scheme~\cite{Hoang:1998ng, Hoang:1998hm, Hoang:1999ye}, which has been
tested in the calculations of numerous observables.  (Using the $\ov{\rm MS}$
mass yields a poorly behaved perturbation series, for the reasons mentioned at
the end of Sec.~\ref{ffsubsec}. Other possible short-distance mass schemes include the PS mass~\cite{Beneke:1998rk} or the kinetic mass~\cite{Czarnecki:1997sz}.) 

The $1S$ scheme defines $m_b^{1S}$ as half of the perturbatively computed
$\Upsilon(1S)$ mass.  It is related to the pole mass as $m_b^{1S} = m_b\, (1 -
2\aS^2/9 + \ldots)$~\cite{Hoang:1998ng, Hoang:1998hm, Hoang:1999ye}, so that we may treat the pole mass as the function $m_b(m_b^{1S}) = m_b^{1S}(1 +
2\aS^2/9 + \ldots)$. Neglecting
higher order terms, as done throughout this paper, is a good approximation in
all cases where they are known, including the evaluation
of~$R_{1,2}$~\cite{Grinstein:2001yg}. We adopt the inputs~\cite{Ligeti:2014kia},
\begin{equation}
	\label{eq:mbdmbc}
	m_b^{1S} = (4.71 \pm 0.05)\,\GeV\,, \qquad \delta m_{bc} = m_b - m_c = (3.40 \pm 0.02)\,\GeV\,,
\end{equation}
from fits to inclusive $B\to X_c l \bar\nu$ spectra and other determinations of
$m_b^{1S}$. We eliminate $m_c$ using $ m_c = m_b(m_b^{1S}) -  \delta m_{bc}$,
and extract $\bar\Lambda$ via
\begin{equation}
\bar\Lambda = \overline{m}_B - m_b(m_b^{1S}) + \lambda_1/(2m_b^{1S})\,.
\end{equation}
Here $\ov m_B = (m_B + 3m_{B^*})/4 \simeq 5.313$\,GeV is the spin-averaged meson
mass, and we use $\lambda_1 = -0.3$\,GeV$^2$~\cite{Ligeti:2014kia}. Enforcing the 
cancellation of the leading renormalon
is equivalent to using $m_b(m_b^{1S}) \to m_b^{1S}$ everywhere in Eqs.~\eqref{eqn:BD1m} and 
\eqref{eqn:BDs1m}, except in the $\bar\Lambda/m_{c,b}$ terms that are 
not multiplied by subleading Isgur-Wise functions.

We match the QCD and HQET theories at scale $\mu^2 = m_b m_c$, corresponding to
$\aS \simeq 0.26$. The $1S$ scheme then yields, for example, the following SM
predictions for $R_{1,2}(1)$
\begin{align}\label{R121}
	R_1(1) &\simeq 1.34 - 0.12\, \eta(1)\,, \nn\\
	R_2(1) &\simeq 0.98 - 0.42\, \eta(1) - 0.54\, \hat\chi_2(1) \,.
\end{align}
For $R'_{1,2}(1)$ we obtain
\begin{align}\label{R12p1}
  R'_1(1) &\simeq -0.15 + 0.06\, \eta(1) - 0.12\, \eta'(1) \,, \nn\\
  R'_2(1) &\simeq 0.01 - 0.54\, \hat\chi'_2(1)  + 0.21\,\eta(1) - 0.42\, \eta'(1)\,.
\end{align}
For completeness, the similar relations for $R_{0,3}$ are
\begin{align}\label{R031}
	R_3(1) &\simeq 1.19  - 0.26\, \eta(1) - 1.20\, \hat\chi_2(1)\,,\nn\\
	R_0(1) &\simeq 1.09 + 0.25\, \eta(1)\,, \nn\\
	R'_3(1) &\simeq -0.08 - 1.20\, \hat\chi'_2(1) + 0.13 \,\eta(1) - 0.26\, \eta'(1)\,,\nn\\
	R'_0(1) &\simeq -0.18 + 0.87\, \hat{\chi}_2(1) + 0.06\,\eta(1) + 0.25\, \eta'(1)\,.
\end{align}

\section{Combined fit to \texorpdfstring{$B\to D^* l\bar\nu$}{bdln} and \texorpdfstring{$B\to D l\bar\nu$}{bdsln}}
\label{sec:fit}

\subsection{Parametrization of the \texorpdfstring{$w$}{w} dependence}

Unitarity and analyticity provide strong constraints on the shapes of the
$\bddselln$ form factors~\cite{Boyd:1995sq, Boyd:1997kz, Caprini:1997mu,
Boyd:1997qw, Hill:2006ub, Bourrely:2008za}.  It is common to employ a
parametrization of the $\bdelln$ form factor $\mathcal{G}(w)$, defined in
Eq.~\eqref{eqn:GFdef}, via the conformal mapping $\mathpzc{z}(w) = (\sqrt{w+1}
- \sqrt{2})/(\sqrt{w+1} + \sqrt{2})$. Unitarity constraints yield, e.g.,
$\mathcal{G}(w)/\mathcal{G}(1) \simeq 1 - 8 \rho^2 \mathpzc{z} + (51. \rho^2 -
10.)\mathpzc{z}^2 - (252.\rho^2 - 84.)\mathpzc{z}^3$, in which $\rho^2 = -
\mathcal{G}'(1)/\mathcal{G}(1)$ is a slope parameter~\cite{Caprini:1997mu}. The
convergence of this expansion may be optimized by parametrizing it in a way that
minimizes the range of the expansion parameter, via
\begin{equation}
	\mathpzc{z}_*(w) = \frac{\sqrt{w+1} - \sqrt{2}\,a}{\sqrt{w+1} + \sqrt{2}\,a}\,,
	\qquad a = \bigg( \frac{1 + r_D}{2\, \sqrt{r_D}} \bigg)^{1/2}\,. 
\end{equation}
For $\bdln$, $|z_*| \le 0.032$. The unitarity constraints suggest a form factor parametrization of the form
\begin{equation}
	\label{eqn:GUC}
	\frac{\mathcal{G}(w)}{\mathcal{G}(w_0)} \simeq 1 - 8a^2\rho_*^2\mathpzc{z}_* + \big(V_{21} \rho_*^2 - V_{20}\big)\mathpzc{z}_*^2\,.
\end{equation}
Here $w_0 = 2a^2 -1 \simeq 1.28$ is defined such that $\mathpzc{z}_*(w_0) = 0$,
while $V_{21} \simeq 57.$ and $V_{20} \simeq 7.5$ are obtained numerically from
Ref.~\cite{Caprini:1997mu}. The uncertainty in the coefficient of the
$\mathpzc{z}_*^2$ term in Eq.~\eqref{eqn:GUC} may be
sizable~\cite{Caprini:1997mu}. However, the impact of this term on the physical fit results is expected to be small.  

The leading order Isgur-Wise function, $\xi(w)$, may be extracted from the
parametrization in Eq.~\eqref{eqn:GUC} by using Eqs.~\eqref{eqn:BD1m} and
\eqref{eqn:hatHdef}. Keeping terms to $\mathcal{O}(\varepsilon_{c,b}(w-1))$, we
can approximate the subleading Isgur-Wise functions as
\begin{equation}
\label{eqn:FIWp}
	\hat{\chi}_2(w)  \simeq \hat{\chi}_2(1) + \hat{\chi}'_2(1)(w-1)\,,\qquad \hat{\chi}_3(w)  \simeq \hat{\chi}'_3(1)(w-1)\,,\qquad
	\eta(w)  \simeq \eta(1) + \eta'(1)(w-1)\,, 
\end{equation}
since $\hat{\chi}_3(1) = 0$. One finds at
$\mathcal{O}(\varepsilon_{c,b},\, \aS)$, 
\begin{align}\label{eqn:Xopt}
\frac{\xi(w)}{\xi(w_0)}
\simeq 1 & - 8 a^2 \brhosq \mathpzc{z}_* + \mathpzc{z}_*^2\, \bigg\{V_{21}\brhosq - V_{20}
  + (\varepsilon_b - \varepsilon_c)\bigg[2\,\Xi\, \eta'(1)\, \frac{1-\rD}{1+\rD}\bigg]\nn \\
& + (\varepsilon_b + \varepsilon_c)\bigg[\Xi\, \big[12 \hat{\chi}'_3(1)-4 \hat{\chi}_2(1)\big]
  - 16\big[(a^2-1)\,\Xi - 16a^4 \big]\hat{\chi}'_2(1)\bigg] \nn \\ 
& + \haS \bigg[\Xi \bigg(\!C_{V_1}^{\prime}(w_0)+\frac{C_{V_3}(w_0) + \rD C_{V_2}(w_0)}{1+\rD}\bigg)
  + 2a^2 (\Xi-32a^2) \frac{C'_{V_3}(w_0) + \rD C'_{V_2}(w_0)}{1+\rD}\nn \\*
& \qquad - 64a^6\, \frac{ C''_{V_3}(w_0) + \rD C''_{V_2}(w_0)}{1+\rD} 
  - 32 a^4 C''_{V_1}(w_0) \bigg]\bigg\}\,,
\end{align}
where $\Xi = 64 a^4 \brhosq  - 16 a^2 - V_{21}$. The slope parameter 
$\brhosq = -\xi'(w_0)/\xi(w_0)$ is related to the slope $\rho_*^2 =
-\mathcal{G}'(w_0)/\mathcal{G}(w_0)$ via
\begin{align}
	\brhosq - \rho_*^2 
	& = (\varepsilon_b + \varepsilon_c)\big[12\hat{\chi}'_3(1)  - 4\hat{\chi}_2(1) - 16(a^2-1)\hat{\chi}'_2(1)\big] + 2(\varepsilon_b - \varepsilon_c) \eta'(1)\frac{1-\rD}{1+\rD}\nn \\
	& \qquad +\haS \bigg[ \frac{\rD C_{V_2}(w_0) + C_{V_3}(w_0)}{1+ \rD} + C'_{V_1}(w_0)
	 +  2a^2\, \frac{\rD C'_{V_2}(w_0) + C'_{V_3}(w_0)}{1+ \rD}\bigg] \,.
\end{align}
Enforcing $\xi(1) =1$, one may directly extract $\xi(w_0)$ via evaluation of
Eq.~\eqref{eqn:Xopt} at the zero recoil point, $\mathpzc{z}_*(w=1) =
(1-a)/(1+a)$, and thereby obtain a properly normalized parametrization for
$\xi(w)$.  Since $\eta(1)$ does not appear in Eq.~(\ref{eqn:Xopt}), this implies
that constraining $\xi(w)$ in itself does not constrain $\eta(1)$, which is the
largest unknown contribution in $R_{1,2}(1)$.

This expression for $\xi(w)$, combined with the HQET expansions in
Eqs.~\eqref{eqn:BD1m} and \eqref{eqn:BDs1m}, allows one to parametrize all
$\Bbar \to \dds$ form factors in terms of six parameters: $\brhosq$,
$\hat\chi_2(1)$, $\hat\chi'_2(1)$, $\hat\chi'_3(1)$, $\eta(1)$ and $\eta'(1)$.
The normalizations of the form factors are also fixed by Eq.~\eqref{eqn:Xopt},
thus $|V_{cb}|$ may be determined from a global fit to overall rates without
using lattice results.

\subsection{QCD sum rule inputs}

The subleading Isgur-Wise functions have only been calculated using model
dependent methods, and are not yet available from lattice QCD.  The two-loop QCD
sum rule (QCDSR) calculations~\cite{Ligeti:1993hw, Neubert:1992wq,
Neubert:1992pn} imply that the subleading Isgur-Wise function $\eta(w)$ is
approximately constant. The functions $\hat\chi_{2,3}$, which parametrize
corrections from the chromomagnetic term in the subleading HQET Lagrangian, are
small, in agreement with quark model intuition.

The QCD sum rule results are obtained at a fixed scale.  The scale dependence
can be removed from $\hat\chi_{2,3}$ by defining ``renormalization improved"
functions, $\hat\chi^{\text{ren}}_{2,3}$~\cite{Neubert:1993mb}. These are
obtained by multiplying the results of Refs.~\cite{Neubert:1992wq,
Neubert:1992pn} for $\hat\chi_{2,3}$ by $[\aS(\Lambda)]^{3/\beta_0} \sim 1.4$, 
where $\Lambda \sim 1\,\GeV$ and $\beta_0 = 9$ for three light flavors. For
these renormalized subleading Isgur-Wise functions, we use
\begin{gather}
\hat{\chi}^{\text{ren}}_2(1) = -0.06 \pm 0.02\,, \qquad
 	 \hat{\chi}^{\prime\,\text{ren}}_2(1) = 0 \pm 0.02\,, \qquad
	\hat{\chi}^{\prime\,\text{ren}}_3(1) = 0.04 \pm 0.02\,, \nn\\*
\eta(1)  = 0.62 \pm 0.2\,,  \qquad  \eta'(1) = 0 \pm 0.2\,. 
\label{eqn:qcdsr}
\end{gather}
These central values reproduce $\hat{L}_{1\ldots 6}$ in
Ref.~\cite{Caprini:1997mu}, often used to predict
$R_{1,2}$ and $R(D^{(*)})$.

We assign relatively large uncertainties, to permit assessment of possible pulls
of the experimental data from these QCDSR predictions. Replacing
$\hat\chi_{2,3}$ with $\hat\chi_{2,3}^{\text{ren}}$, the Wilson coefficient of
the chromomagnetic operator receives a corresponding $\aS(\mu)^{3/\beta_0}$
factor at the matching scale $\mu = \sqrt{m_b\, m_c}$, partly canceling the
above $[\aS(\Lambda)]^{3/\beta_0}$ enhancement. For ease of comparison with
the literature we ignore this, as it can be viewed as a higher order correction,
and is in any case covered by the large assigned uncertainties. We ignore
correlations in the QCDSR results (arising from the common calculational
method), which is conservative.  

Using Eqs.~\eqref{eqn:qcdsr} in Eqs.~\eqref{eqn:R12exp} yield expressions for
$R_{1,2}(w)$ as polynomials in $(w-1)$, with the coefficients and their
uncertainties correlated by HQET.  In Ref.~\cite{Caprini:1997mu}, the central
values in Eq.~\eqref{eqn:qcdsr} were used to write $R_{1,2}(w)$ as quadratic
polynomials, without quoting any theory uncertainties on their slopes and
curvatures. It subsequently become standard practice in experimental $|V_{cb}|$
and $R_{1,2}$ measurements to fit for $R_{1,2}(1)$, while fixing $R'_{1,2}(1)$
and $R''_{1,2}(1)$ to their quoted central values~\cite{Caprini:1997mu}.  Such
an approach is inconsistent with the simultaneous use of the HQET constraints
and the QCDSR results.  For example, the present world average central value,
$R_1(1) \simeq 1.4$, cannot simultaneously satisfy the HQET prediction for
$R_1(1)$ in Eq.~\eqref{R121} and the QCDSR expectation $\eta(1) > 0$, which
holds at the $3\sigma$ level, and is used elsewhere in the same fit.  A
consistent treatment of these form factor ratios is absent from the derivations
of the state-of-the-art predictions for $R(D^{(*)})$ in the SM (except for LQCD
$R(D)$ predictions) and in the presence of new physics~\cite{Fajfer:2012vx,
Tanaka:2012nw}.

We now proceed to assess the importance of obeying the HQET relations between
different form factors, and of including the uncertainties in the QCDSR
predictions in Eq.~\eqref{eqn:qcdsr}. These effects will be important in the
future, to systematically improve the SM predictions.

\subsection{Fit scenarios}
A simultaneous fit of the six parameters $\brhosq$, $\hat\chi_2(1)$,
$\hat\chi'_2(1)$, $\hat\chi'_3(1)$, $\eta(1)$, and $\eta'(1)$ to the $\bddsln$
rates can be carried out with the present data. Such a fit fixes both the shapes
and normalizations of the $\bddsln$ rates, without any theory input other than
the HQET expansion. However, one expects large uncertainties at present, because
of the limited experimental precision and the number of subleading HQET
parameters. One may instead use QCD sum rule predictions and/or lattice QCD
results to constrain the fit, increasing sensitivity to $\brhosq$. The fit
propagates the uncertainties on the subleading Isgur-Wise functions into the fit
result, and allows the data to further constrain the subleading contributions.

Our fit relies on the HQET predictions and unitarity constraints to determine
the ratios and shapes of the form factors.  The form factors at zero recoil,
$\mathcal{G}(1)$ and $\mathcal{F}(1)$, have been computed in lattice QCD (LQCD),
providing state-of-the-art predictions for the normalizations of the $\bddsln$
rates.  The most precise lattice QCD predictions at zero recoil
are~\cite{Bailey:2014tva, Lattice:2015rga}
\begin{equation}\label{eq:F1}
	\mathcal{G}(1)_{\text{LQCD}} = 1.054(8)\,, \qquad 
	\mathcal{F}(1)_{\text{LQCD}} = 0.906(13)\, ,
\end{equation}
where we combined the quoted systematic and statistical uncertainties.  Although
these normalizations may be expected to drop out of the predictions for
$R(\dds)$, they do influence the fit to the differential decay distributions and
hence the resulting form factor ratios.  Making use of these lattice constraints
leads to our first fitting scenario:
\begin{itemize}
\item[\scriptsize$\blacksquare$] Rescale the $\Bbar \to D$ and $\Bbar \to D^*$
form factors in the fit by $\mathcal{G}(1)_{\text{LQCD}}/\mathcal{G}(1)$ and
$\mathcal{F}(1)_{\text{LQCD}}/\mathcal{F}(1)$, respectively, such that the rates
at $w=1$ agree with the lattice predictions. We refer to this fit as
``$\lqcdGF$".
\end{itemize}
Measurements of the rate normalizations are, however, subject to relatively
large systematic uncertainties. For example, the calibration of the hadronic tagging
efficiency produces systematic uncertainties of the order of a few
percent~\cite{Abdesselam:2017kjf}. To compare the best-fit shapes without
lattice constraints and such systematic effects, we consider a second scenario:
\begin{itemize}
\item[\scriptsize$\blacksquare$] Allow the normalizations of the $\bdln$ and
$\bdsln$ rates to float independently. This approach only uses $\bddsln$ shape
information to constrain the form factors, but no theory input for the
normalizations at zero-recoil, and is independent of lattice information. 
We refer to this fit as ``$\normDDs$".
\end{itemize} 
For each fit, we apply (relax) the QCDSR constraints, exploring a ``constrained"
(``unconstrained") fit. The QCDSR constrained fits are denoted with a suffix
``$+\text{SR}$". Both $\lqcdGF$ and $\normDDs$ fits alter the overall
normalizations the $\bdln$ and $\bdsln$ rates, but leave the HQET expansions of
the form factors unchanged. Thus, they can be considered as introducing an extra
source of heavy quark symmetry breaking in the normalizations (to effectively
account for higher order effects), while still preserving the form factor
relations independently in Eqs.~\eqref{eqn:BD1m} and \eqref{eqn:BDs1m}.

Since lattice QCD predictions are also available for $w \ge 1$ for the $\bdln$
form factors $f_+(w)$ and $f_0(w)$, it is possible to obtain a prediction
for the slope parameter, $\brhosq$, from them. This leads to a third fit
approach, namely:
\begin{itemize}
\item[\scriptsize$\blacksquare$] Extract $\xi(w)$, including the slope parameter
$\brhosq$, by fitting to the $w\geq 1$ lattice QCD data for $B \to D$, and apply it simultaneously
with the LQCD normalization of $B \to D^*$ at $w=1$. We refer to this fit as ``$\lqcdXw$".
\end{itemize}
In a ``theory only" version of this fit, denoted by ``$\lqcdXwQth$", one fully
constrains the $\bddsln$ differential rates without any experimental input; the
only fit is to lattice data and QCDSR constraints. For the ``$\lqcdXwQ$" fit, we
combine the $w\ge1$ $B \to D$ and $w=1$ $B \to D^*$ lattice data with QCDSR
constraints and the experimental information, to include all available
information and explore possible tensions.  We summarize the inputs of the
various fit scenarios pursued in this paper in Table~\ref{tab:FitKey}.

\begin{table}[t]
\newcolumntype{C}{ >{\centering\arraybackslash } m{2cm} <{}}
\begin{tabular}{l|C|ccc|C}
	\hline
	\multicolumn{1}{c|}{\multirow{2}{*}{Fit}}  &  \multirow{2}{*}{QCDSR}
  	&  \multicolumn{3}{c|}{Lattice QCD}  &  \multirow{2}{*}{Belle Data}\\
	&  &  $\mathcal{F}(1)$  &  $f_{+,0}(1)$ &  $f_{+,0}(w > 1)$  & \\ \hline\hline
	$\lqcdGF$ 	&  ---  &  $\checkmark$	 &  $\checkmark$  &  ---  &  $\checkmark$\\
	$\lqcdGFQ$	&  $\checkmark$  &  $\checkmark$  &  $\checkmark$  &  ---  &  $\checkmark$ \\ \hline
	$\normDDs$ 	&  ---  &  ---  &  ---  &  ---  &  $\checkmark$\\
	$\normDDsQ$ 	&  $\checkmark$  &  ---  &  ---  &  ---  &  $\checkmark$\\ \hline
	$\lqcdXw$ 	&  --- &  $\checkmark$  &  $\checkmark$  &  $\checkmark$  &  $\checkmark$  \\
	$\lqcdXwQ$  	&  $\checkmark$  &  $\checkmark$  &  $\checkmark$  &  $\checkmark$ & $\checkmark$ \\ \hline
	$\lqcdXwQth$ 	&  $\checkmark$  &  $\checkmark$  &  $\checkmark$  &  $\checkmark$  & ---\\	
	\hline\hline
\end{tabular}
\caption{Summary of theory and data inputs for each fit scenario.  All use the
HQET predictions to order $\mathcal{O}(\lqcd/m_{c,b})$ and $\mathcal{O}(\aS)$, as well as the
unitarity constraints.}
\label{tab:FitKey}
\end{table}

All fits explored in this paper use the unitarity constraints.  
The consequences of relaxing the unitarity constraints between the slope and the
curvature terms in Eq.~\eqref{eqn:GUC} will be explored in detail
elsewhere~\cite{Vcb:2017}.

\subsection{Data and fit details}

To determine the leading and subleading Isgur-Wise functions and $| V_{cb}|$, we
carry out a simultaneous fit of the available $\bddsln$ spectra.  There are only
two measurements~\cite{Glattauer:2015teq, Abdesselam:2017kjf} which provide
kinematic distributions fully corrected for detector effects.  The measured
recoil and decay angle distributions are analyzed simultaneously by constructing
a standard $\chi^2$ function. Common uncertainties (tagging efficiency,
reconstruction efficiencies, number of $B$-meson pairs) should be treated as
fully correlated between the two measurements and we construct a covariance
using Table~IV in Ref.~\cite{Glattauer:2015teq} and Table~IV in
Ref.~\cite{Abdesselam:2017kjf}.  While Ref.~\cite{Glattauer:2015teq} provides a
full breakdown of the total uncertainty for each measured $w$ bin,
Ref.~\cite{Abdesselam:2017kjf} only provides a breakdown for the total branching
fraction. To construct the desired covariance between both measurements, we
thus assume that there is no shape dependence on the tagging and reconstruction
efficiency uncertainty of Ref.~\cite{Abdesselam:2017kjf}. Comparing this with
the mild dependence on these error sources in Ref.~\cite{Glattauer:2015teq},
this seems a fair approximation of the actual covariance. To take into account
the uncertainties of $m^{1S}_b$ and $\delta m_{bc}$, we introduce both as
nuisance parameters into the fit, assuming Gaussian constraints with
uncertainties given in Eq.~\eqref{eq:mbdmbc}. The $\chi^2$ function is
numerically minimized and uncertainties are evaluated using the usual asymptotic
approximations by scanning the $\Delta \chi^2 = \chi^2_{\rm scan} - \chi^2_{\rm
min}$ contour to find the $+1$ crossing point, which provides the 68\%
confidence level. The constraints from lattice QCD predictions and/or QCD sum
rules are incorporated into the fit assuming (multivariate) Gaussian errors and
are added to the $\chi^2$ function. 

The full fit results are shown in Table~\ref{tab:res}. The ``$\lqcdGF$"
unconstrained fit, i.e., using only the lattice normalizations at $w=1$, yields
\begin{equation}
 	| V_{cb} | = ( 38.8 \pm 1.2 ) \times 10^{-3} \, ,
\end{equation}
to be compared with the current world average~\cite{PDG} 
$|V_{cb}| = (42.2 \pm 0.8) \times 10^{-3}$ and \mbox{$|V_{cb}| = (39.2 \pm 0.7) \times 10^{-3}$}, 
from inclusive and exclusive $b \to c\, l \bar \nu_l$ decays,
respectively. The uncertainties of the subleading Isgur-Wise parameters are
sizable.  There is no sensitivity to  disentangle $\eta'(1)$ from $\brhosq$,
so we fix $\eta'(1)$ to be zero for all QCDSR unconstrained fits.
Including the QCDSR constraints in the ``$\lqcdGFQ$" fit yields
\begin{equation}
 	| V_{cb}| = ( 38.5 \pm 1.1) \times 10^{-3} \, ,
\end{equation}
resulting in almost the same $|V_{cb}|$ value. The normalization of $\eta(1)$ is
comparable between these two fits, at about half the value of the QCDSR
expectation.  Both fits have reasonable $\chi^2$ values, corresponding to fit
probabilities of 64\% each.

\begin{table}[t!]
\newcolumntype{C}{ >{\centering\arraybackslash $} m{2.25cm} <{$}}
\newcolumntype{D}{ >{\centering\arraybackslash $} m{1.75cm} <{$}}
\resizebox{\linewidth}{!}{
\begin{tabular}{D|CC|CC|CC|C}
	\hline
	& \lqcdGF & \lqcdGFQ & \normDDs & \normDDsQ & \lqcdXw & \lqcdXwQ & \lqcdXwQth\\
	\hline
	\hline
	\chi^2	&	40.2	&	44.0	&	38.7	&	43.1	&	49.0	&	53.8	&	7.4	\\
\text{dof}	&	44	&	48	&	43	&	47	&	48	&	52	&	4	\\ \hline
|V_{cb}| \times 10^3	&	38.8 \pm 1.2	&	38.5 \pm 1.1	&	\text{---}	&	\text{---}	&	39.1 \pm 1.1	&	39.3 \pm 1.0	&	\text{---}	\\
\mathcal{G}(1)	&	1.055 \pm 0.008	&	1.056 \pm 0.008	&	\text{---}	&	\text{---}	&	1.060 \pm 0.008	&	1.061 \pm 0.007	&	1.052 \pm 0.008	\\
\mathcal{F}(1)	&	0.904 \pm 0.012	&	0.901 \pm 0.011	&	\text{---}	&	\text{---}	&	0.898 \pm 0.012	&	0.895 \pm 0.011	&	0.906 \pm 0.013	\\
\brhosq	&	1.17 \pm 0.12	&	1.19 \pm 0.07	&	1.06 \pm 0.15	&	1.19 \pm 0.08	&	1.33 \pm 0.11	&	1.24 \pm 0.06	&	1.24 \pm 0.08	\\
\hat\chi_2(1)	&	-0.26 \pm 0.26	&	-0.07 \pm 0.02	&	0.36 \pm 0.62	&	-0.06 \pm 0.02	&	0.13 \pm 0.22	&	-0.06 \pm 0.02	&	-0.06 \pm 0.02	\\
\hat\chi_2'(1)	&	0.21 \pm 0.38	&	-0.00 \pm 0.02	&	0.14 \pm 0.39	&	-0.00 \pm 0.02	&	-0.36 \pm 0.28	&	-0.00 \pm 0.02	&	-0.00 \pm 0.02	\\
\hat\chi_3'(1)	&	0.02 \pm 0.07	&	0.05 \pm 0.02	&	0.18 \pm 0.19	&	0.04 \pm 0.02	&	0.09 \pm 0.07	&	0.05 \pm 0.02	&	0.04 \pm 0.02	\\
\eta(1)	&	0.30 \pm 0.04	&	0.30 \pm 0.03	&	-0.56 \pm 0.80	&	0.35 \pm 0.14	&	0.30 \pm 0.04	&	0.30 \pm 0.03	&	0.31 \pm 0.04	\\
\eta'(1)	&	0~\text{(fixed)}	&	-0.12 \pm 0.16	&	0~\text{(fixed)}	&	-0.11 \pm 0.18	&	0~\text{(fixed)}	&	-0.05 \pm 0.09	&	0.05 \pm 0.10	\\
m_b^{1S}~\text{[GeV]}	&	4.70 \pm 0.05	&	4.70 \pm 0.05	&	4.71 \pm 0.05	&	4.70 \pm 0.05	&	4.71 \pm 0.05	&	4.71 \pm 0.05	&	4.71 \pm 0.05	\\
\delta m_{bc}~\text{[GeV]}	&	3.40 \pm 0.02	&	3.40 \pm 0.02	&	3.40 \pm 0.02	&	3.40 \pm 0.02	&	3.40 \pm 0.02	&	3.40 \pm 0.02	&	3.40 \pm 0.02	\\
	\hline
	\hline
\end{tabular}}
\caption{Summary of the results for the fit scenarios considered. 
The correlations are shown in Appendix~\ref{app:DC}.}
\label{tab:res}
\end{table}

Neglecting all subleading $\lqcd/m_{c,b}$ contributions in the ``$\lqcdGF$" fit
results in a poorer overall $\chi^2$.  The value of $|V_{cb}|$ decreases
slightly, $|V_{cb}| = (38.2 \pm 1.1) \times 10^{-3}$, with $\chi^2 = 62.6$ for
48 dof, corresponding to a fit probability of 8\%, which is still an acceptable
fit. The slope parameter becomes $\brhosq = 0.93 \pm 0.05$, below those obtained
including the $\lqcd/m_{c,b}$ corrections. The uncertainty of
$\brhosq$ is noticeably smaller due to the smaller number of degrees of freedom
in this fit. The value of $| V_{cb} |$ is only weakly affected by this 
shift in $\brhosq$. 

In the ``$\normDDs$" fits, using no LQCD inputs, we use only shape information
to disentangle $\brhosq$ from the subleading contributions, while allowing the
$\bdln$ and $\bdsln$ channels to each have arbitrary normalizations (these fits
cannot determine $|V_{cb}|$).  This results in large uncertainties in the QCDSR
unconstrained fit. Again, $\eta'(1)$ and $\brhosq$ are strongly correlated, so
the former is fixed at zero.  Including the QCDSR constraints in the
``$\normDDsQ$" fit yields results close to those in the ``$\lqcdGFQ$" fit. 

\begin{figure}[t]
\includegraphics[width=0.45\textwidth]{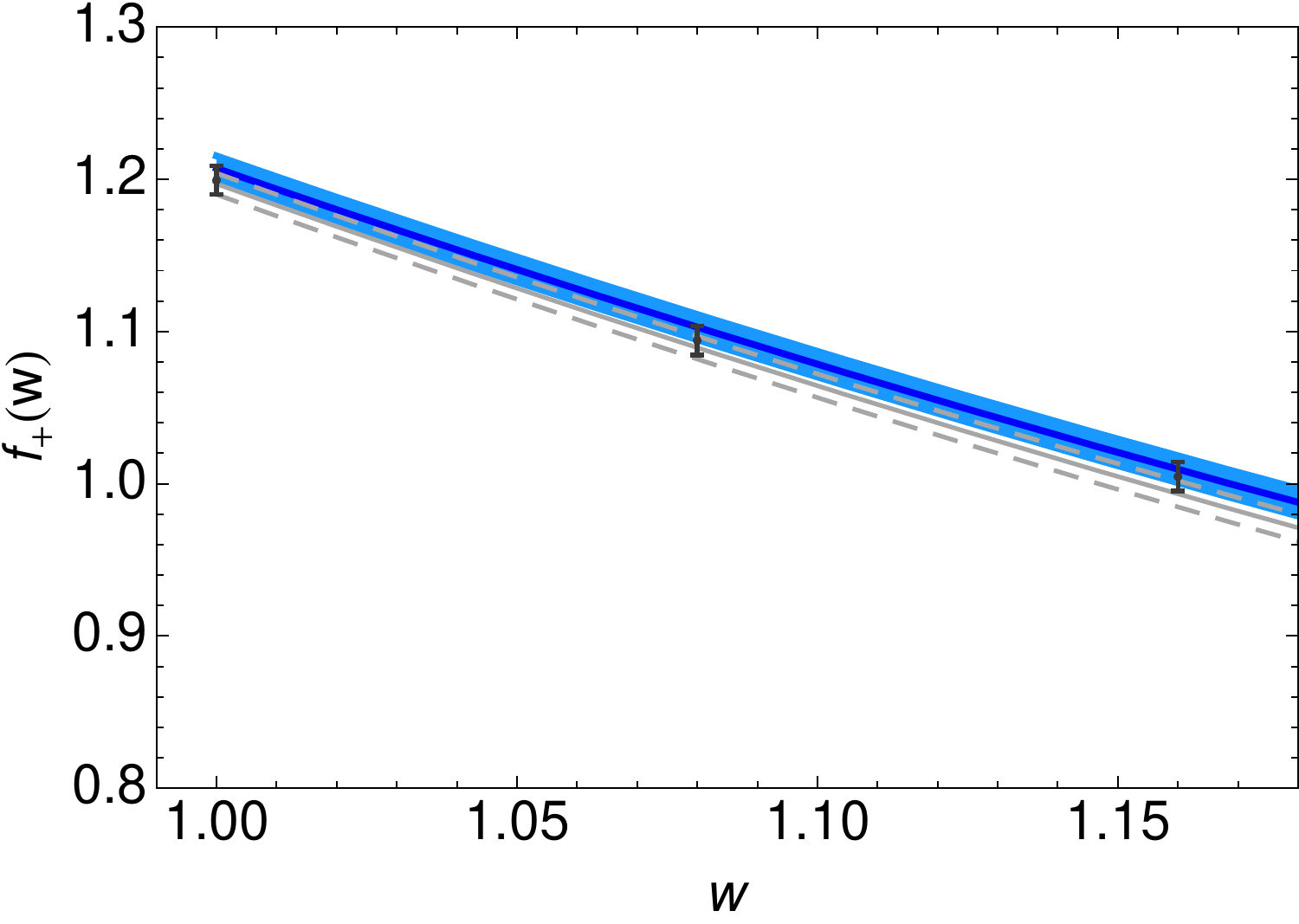} \hspace*{1cm}
\includegraphics[width=0.45\textwidth]{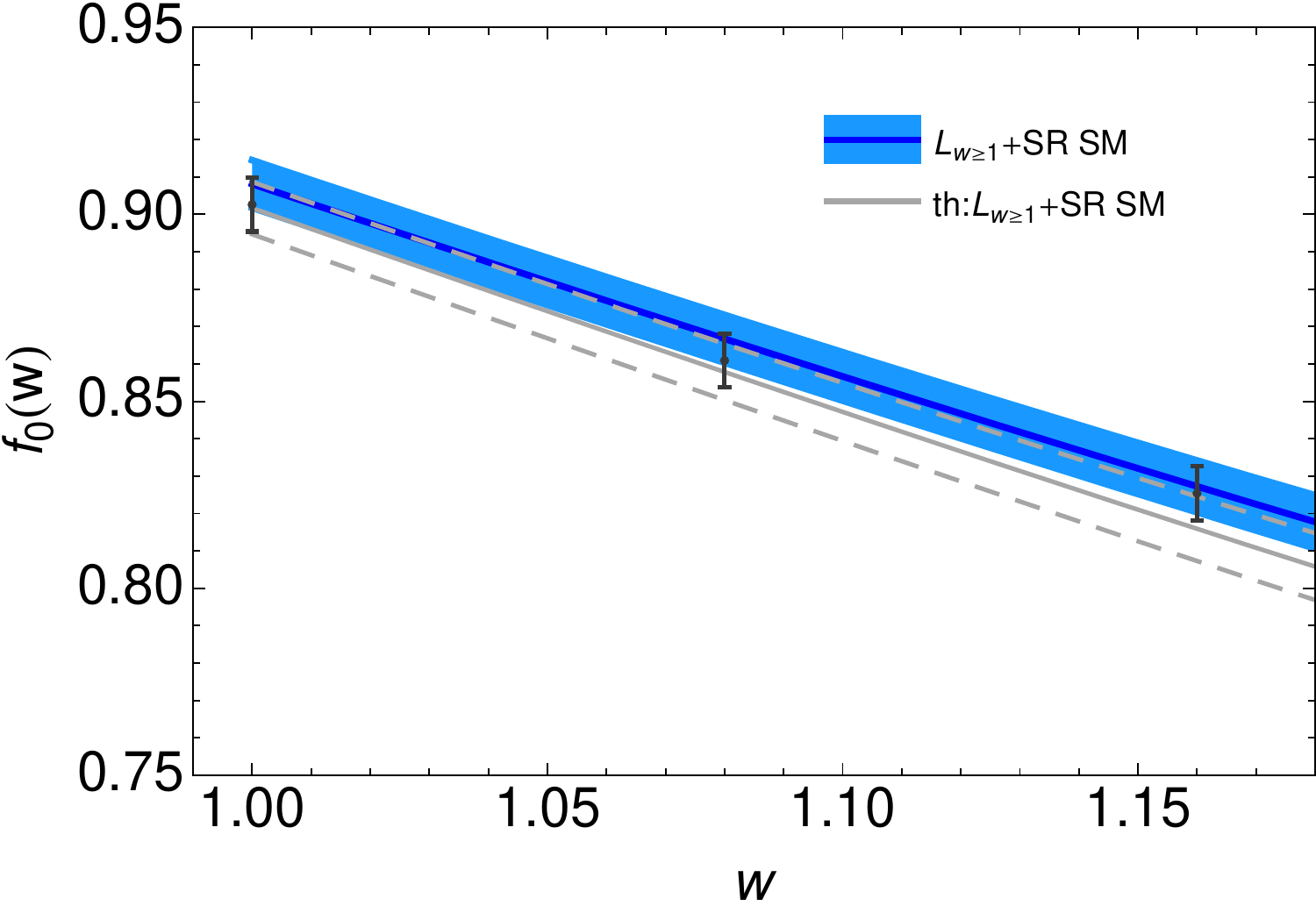}
\vspace*{-10pt}
\caption{The ``$\lqcdXwQth$" fit of the form factors $f_{+,0}$ to the lattice
points listed in Table~\ref{tab:FNALfpf0} is shown (gray solid line). The dashed
gray lines correspond to the 68\% errors. The dark blue line shows the $f_{+,0}$
best fit for ``$\lqcdXwQ$", using lattice points, experimental information, and
QCDSR constraints. The blue band displays the corresponding 68\% CL of this
fit.}
\label{fig:lattice}
\end{figure}

In the ``$\lqcdXwQth$" scenario, which uses no experimental data, fitting the parametrized $\xi(w)$ to the six
lattice points for $f_{+,0}(w)$ in Table~\ref{tab:FNALfpf0} and
$\mathcal{F}(1)$ in Eq.~\eqref{eq:F1}, results in a slope parameter
\begin{equation}
	\brhosq = 1.24 \pm 0.08 \,.
\end{equation}
The fitted $w$ spectra are shown in Fig.~\ref{fig:lattice} (gray curves), together with the
lattice data points. The $\chi^2$ of the fit is 7.4, corresponding to a fit
probability of 11\% with $7-3=4$ degrees of freedom. The value for the slope is
in good agreement with the slope obtained from the QCDSR constrained and
unconstrained ``$\lqcdGF$" and ``$\normDDs$" fits.

\begin{table}[t!]
\begin{tabular}{c|ccc}
\hline\hline
  Form factor & $w = 1.0$ & $w = 1.08$ & $w = 1.16$ \\ \hline
  $f_{+}$ & $1.1994 \pm 0.0095$  &  $1.0941 \pm 0.0104$  &  $1.0047 \pm 0.0123$ \\
  $f_{0}$ & $0.9026 \pm 0.0072$  &  $0.8609 \pm 0.0077$  &  $0.8254 \pm 0.0094$ \\
\hline\hline
\end{tabular}
\caption{The predictions for the form factors $f_{+,0}$ at $w =
1.0,\,1.08,\,1.16$ using the synthetic data results of Ref.~\cite{Lattice:2015rga}.
The correlations can be found in Table VII in Ref.~\cite{Lattice:2015rga}.}
\label{tab:FNALfpf0}
\end{table}

\begin{figure}[t]
\includegraphics[width=0.45\textwidth]{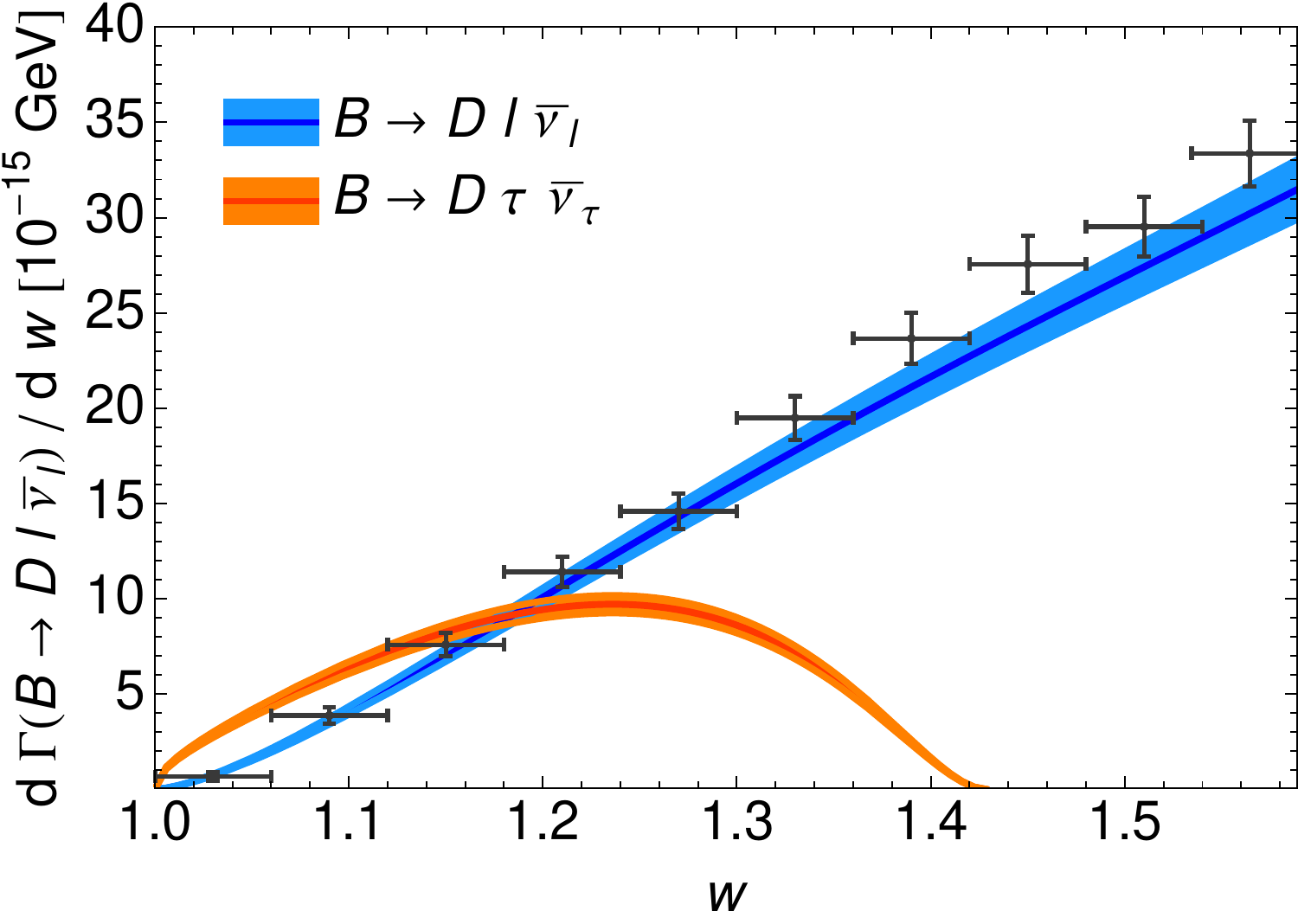}\\
\includegraphics[width=0.45\textwidth]{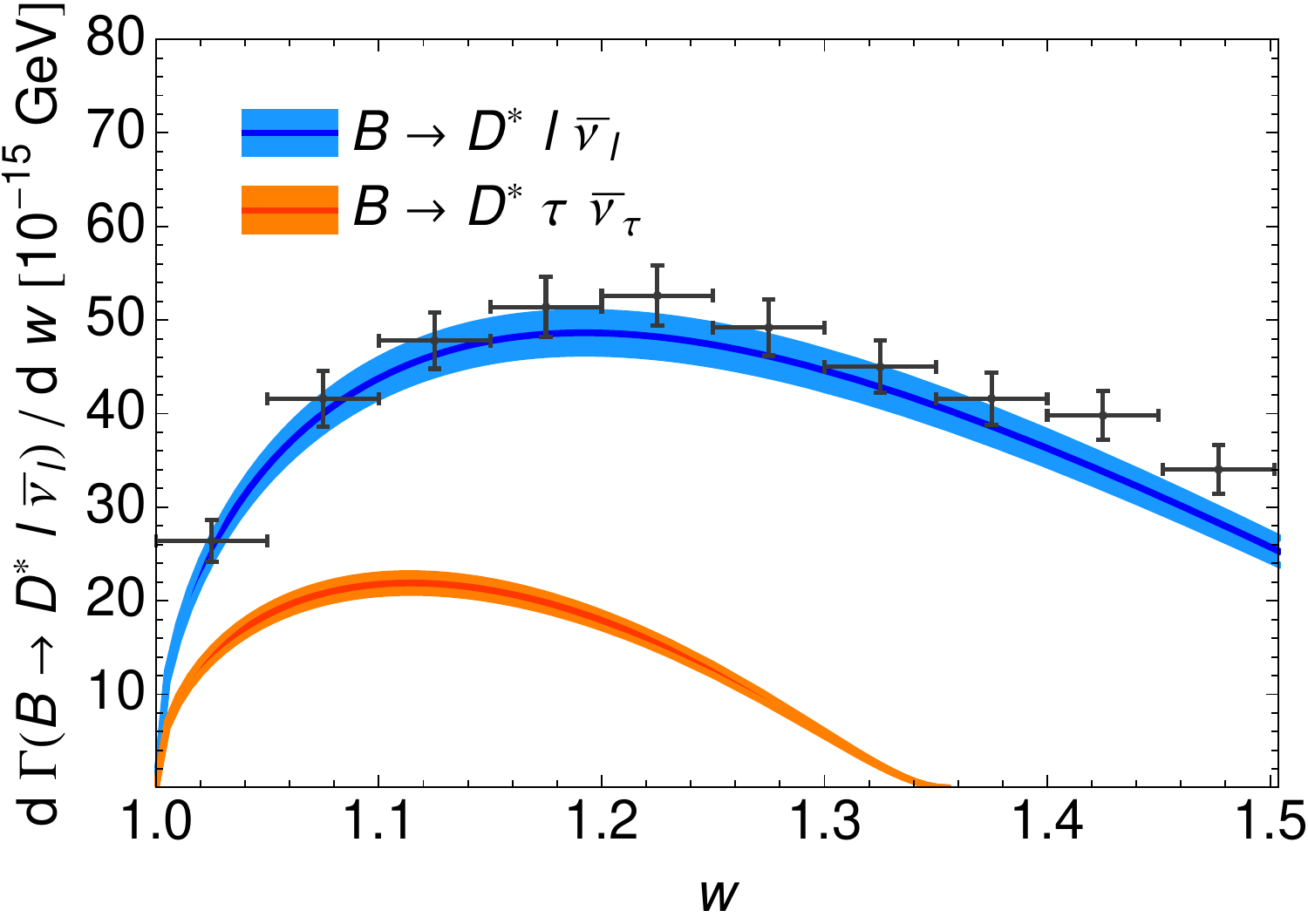} \hspace*{1cm}
\includegraphics[width=0.45\textwidth]{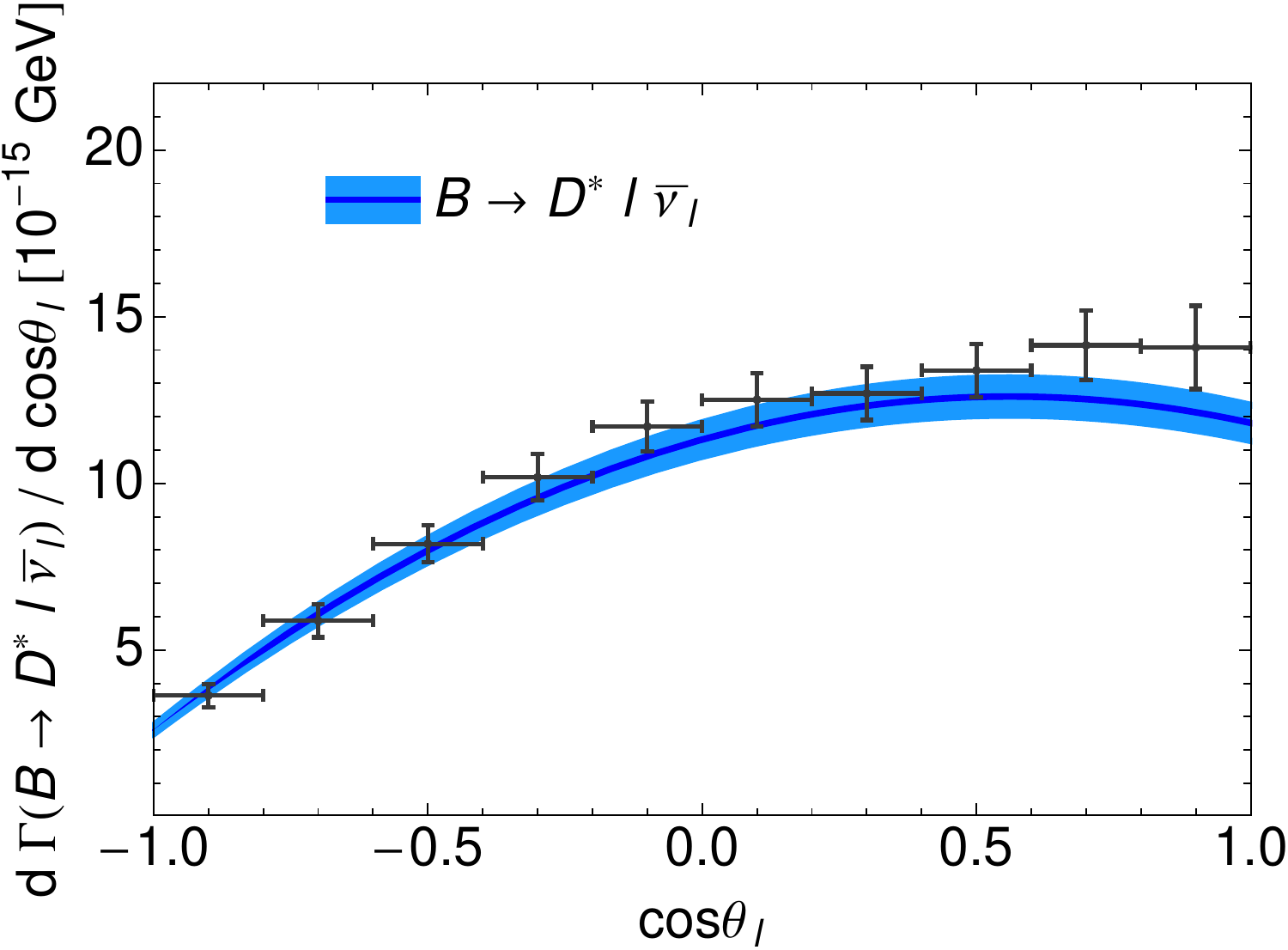} \\
\includegraphics[width=0.45\textwidth]{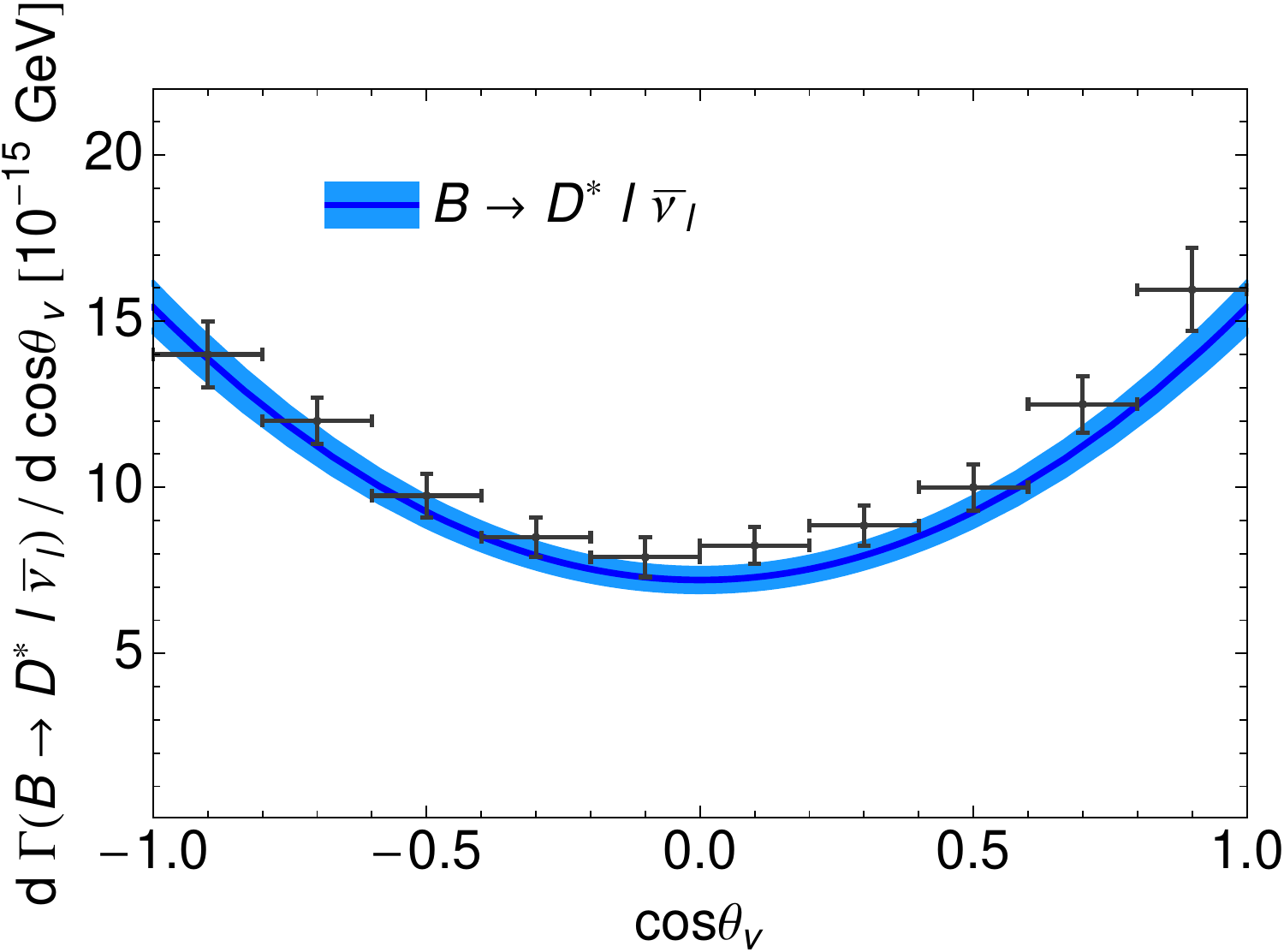} \hspace*{1cm}
\includegraphics[width=0.45\textwidth]{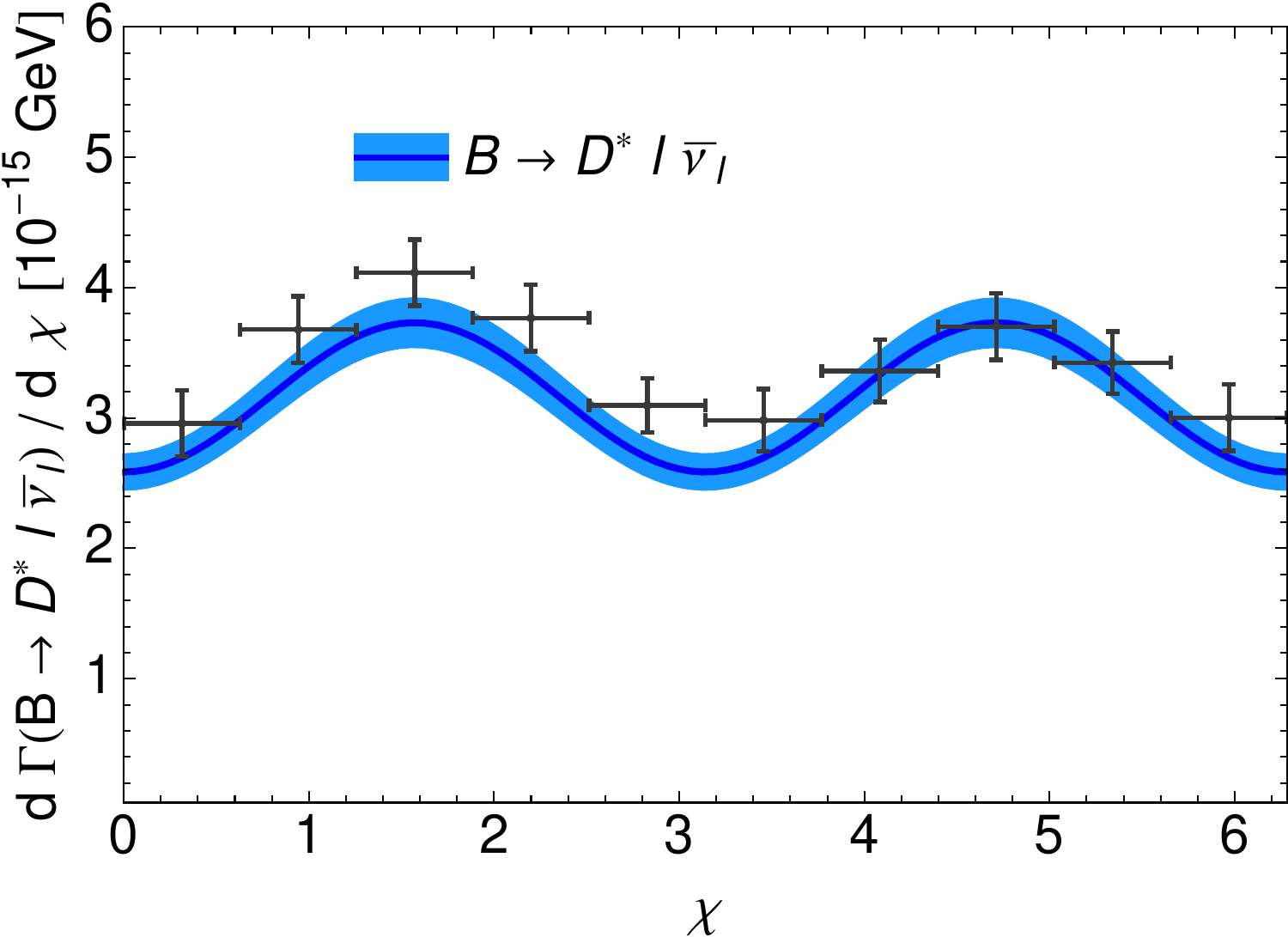} \\
\vspace*{-10pt}
\caption{The measured $\bddsln$ decay distributions~\cite{Glattauer:2015teq,
Abdesselam:2017kjf} compared to the best fit contours (dark blue curves) for the
``$\lqcdXwQ$" fit, using LQCD at all $w$ and QCDSR constraints. The blue bands
show the 68\% CL regions.  The orange curves and bands show the central values
and the 68\% CL regions of the fit predictions for $\d\Gamma(\bddstn) / \d w$.}
\label{fit:latticeData}
\end{figure}

In the ``$\lqcdXw$" fit, all six lattice points for $f_{+,0}(w)$ in
Table~\ref{tab:FNALfpf0} and $\mathcal{F}(1)$ in Eq.~\eqref{eq:F1} are fitted
together with the available experimental information. Once again,
$\eta'(1)$ is fixed to zero, as it is strongly correlated with $\brhosq$. The
fit has $\chi^2 = 49$, corresponding to a fit probability of 43\%.  For
$|V_{cb}|$, this fit yields
\begin{equation}
	 |V_{cb}| = (39.1 \pm 1.1) \times 10^{-3}  \, ,
\end{equation}
which is slightly higher than the ``$\lqcdGF$" result. The value of $\brhosq$ is
also higher.

In the ``$\lqcdXwQ$" fit, the QCDSR constraints are included, so that all
 theory and experimental information is incorporated.  The resulting differential $\bddsln$ distributions are shown in
Fig.~\ref{fit:latticeData}, overlaid with the
experimental data, as well as the predictions for the $\bddstn$
differential rates.
The fit has $\chi^2 = 53.8$,
corresponding to a fit probability of 44\%. For $|V_{cb}|$ the fit gives
\begin{equation}
	 |V_{cb}| = (39.3 \pm 1.0) \times 10^{-3}  \,.
\end{equation}
This is higher than the ``$\lqcdGFQ$" result, because the value of $\brhosq$ is
also higher. 

The correlation matrices for all fits are shown in Appendix~\ref{app:DC}. In the
``$\lqcdGF$" and ``$\lqcdXw$" type fits, moderate correlations are seen between
$|V_{cb}|$, $\mathcal{G}(1)$, and $\mathcal{F}(1)$, as expected.  The
correlations are sizable in these fits between $\brhosq$ and the subleading Isgur-Wise
functions.

\begin{figure}[p]
\includegraphics[width=0.45\textwidth]{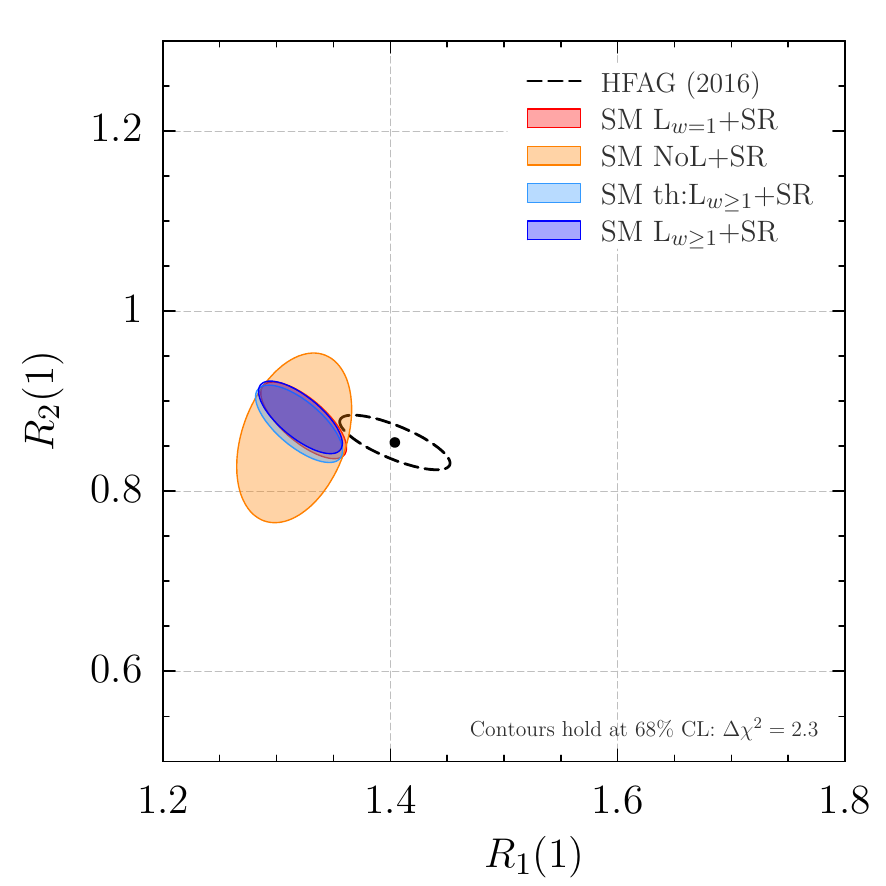} \hspace*{1cm}
\includegraphics[width=0.45\textwidth]{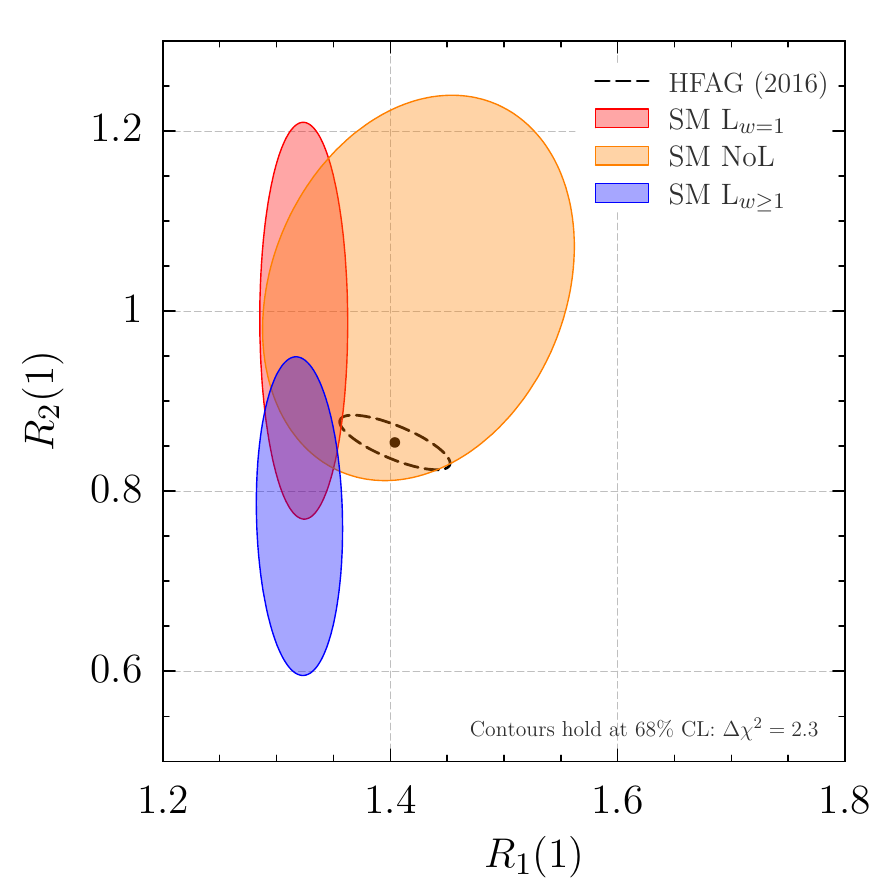}
\caption{The SM predictions for $R_1(1)$ and $R_2(1)$ for the fits imposing
(left) or not imposing (right) the QCDSR constraints in Eq.~(\ref{eqn:qcdsr}). 
The black ellipse shows the world average of the data~\cite{HFAG}. The
fit scenarios are described in the text and in Table~\ref{tab:FitKey}, and the
fit results are shown in Table~\ref{tab:res}.  All contours
correspond to 68\% CL in two dimensions ($\Delta \chi^2 = \chi^2_{\rm scan} -
\chi^2_{\rm min} = 2.3$).}
\label{fit:R1R2}

\vspace*{18pt}
\includegraphics[width=0.45\textwidth]{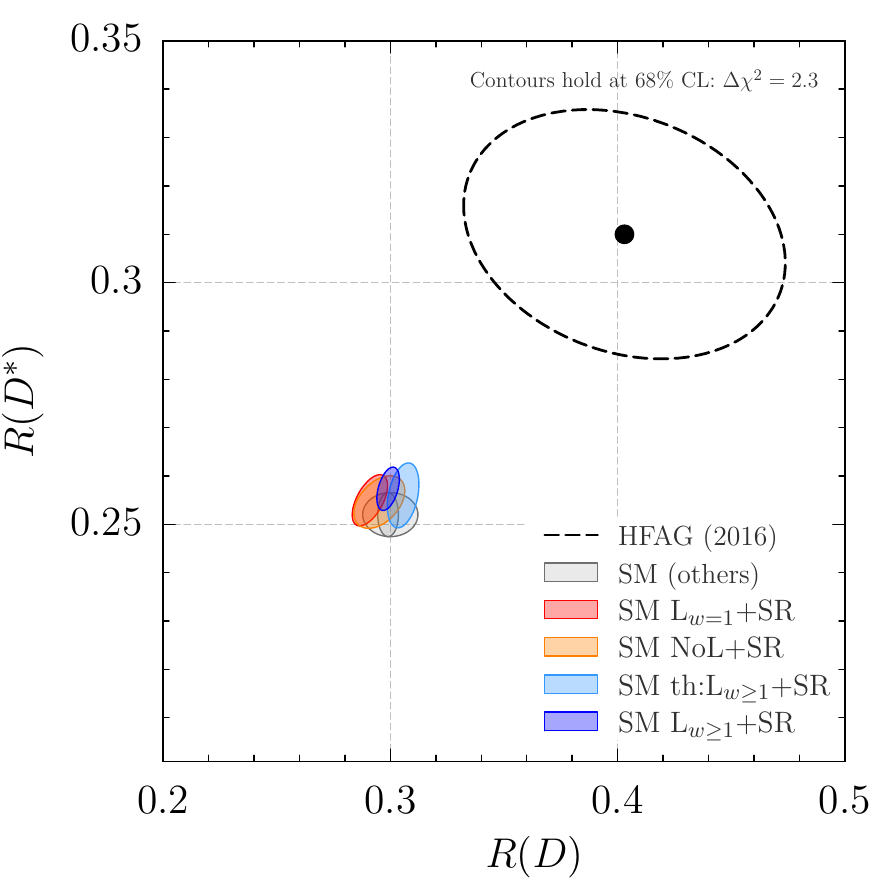} \hspace*{1cm}
\includegraphics[width=0.45\textwidth]{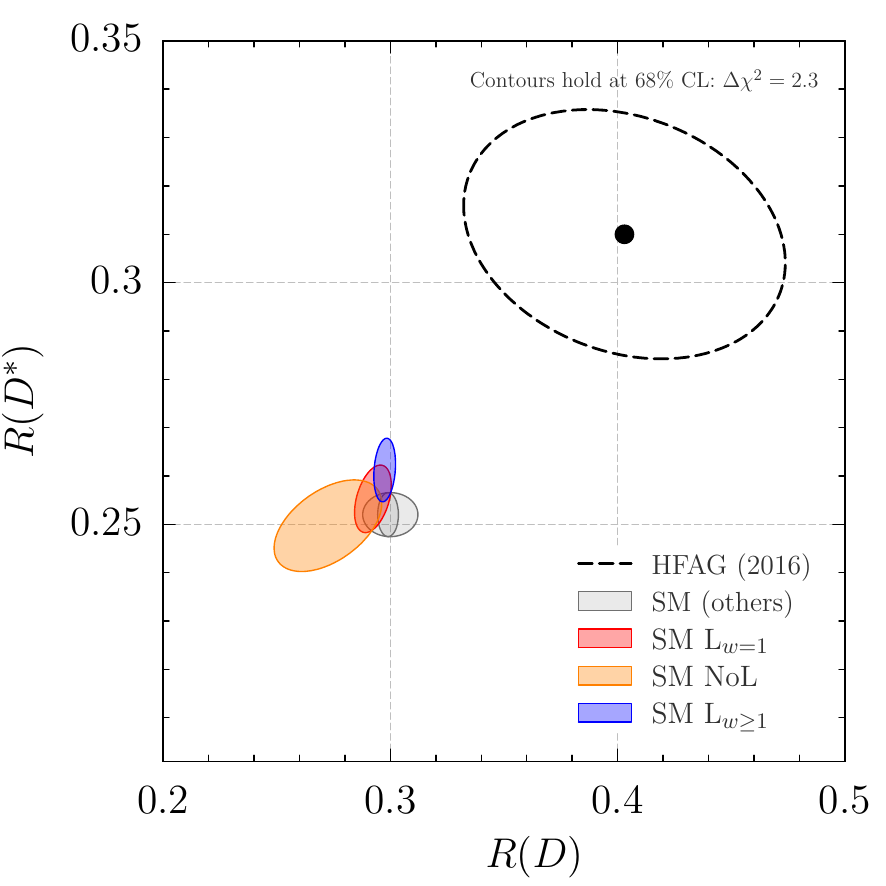}
\caption{The SM predictions for $R(D)$ and $R(D^*)$, imposing (left) or not
imposing (right) the QCDSR constraints (see Table~\ref{tab:RDRDs}).  
Gray ellipses show other SM predictions (last three rows of Table~\ref{tab:RDRDs}). 
The black ellipse shows the world average of the data~\cite{HFAG}. 
 The contours are 68\% CL ($\Delta \chi^2 = 2.3$), hence
the nearly $4\sigma$ tension.}
\label{fit:RDDs}
\end{figure}

A more detailed study of these effects, in particular the extraction of
$|V_{cb}|$, will be presented elsewhere~\cite{Vcb:2017}.  A first comparison
with the CLN parametrization~\cite{Caprini:1997mu}, as implemented by previous
experimental studies, can be done by considering the results for the form
factor ratios $R_1$ and $R_2$, defined in Eq.~\eqref{eqn:R1R2Def}.
Figure~\ref{fit:R1R2} shows the extracted values of $R_{1,2}(1)$ for all fit
scenarios. The results agree with each other and with the
world average of $R_1(1)$ and $R_2(1)$~\cite{HFAG} shown by black ellipses, up to a mild $1\sigma$ tension.
Firm conclusions are difficult to reach, as it is impossible to assess how the
experimental results would change, had the uncertainties in the quadratic
polynomials used to fit $R_{1,2}(w)$ been properly included.  When the QCDSR
constraints are used, the central values satisfy $R_1(1) < 1.34$, as required by
the HQET prediction in Eq.~\eqref{R121} and the constraint $\eta(1) >0$.

\subsection{\texorpdfstring{$R(\dds)$}{Rdds} and new physics }

Using the fitted values for  $\brhosq$, $\hat\chi_2(1)$, $\hat\chi'_2(1)$,
$\hat\chi'_3(1)$, $\eta(1)$, and $\eta'(1)$, one can predict $R(D^{(*)})$ in the
SM and for any new physics four-fermion interaction. Figure~\ref{fit:RDDs} and
Table~\ref{tab:RDRDs} summarize the predicted values of $R(D^{(*)})$ in the SM
for the seven fit scenarios considered.  Our fit results for $R(D)$ are in good
agreement with other predictions in the literature~\cite{Aoki:2016frl,
Bigi:2016mdz}. All our fits using lattice QCD inputs yield $R(D^*)$ above those in
Ref.~\cite{Fajfer:2012vx}. This slightly eases the disagreement with the world average
measurement~\cite{HFAG}. The significance is calculated from $\chi^2$ statistics, taking into
account the full covariance of the theory prediction and the world average
measurement.  The tension between our most precise ``$\lqcdXwQ$" fit and the data is $3.9
\sigma$, with a $p$-value of $11.5 \times 10^{-5}$, to be compared with $8.3
\times 10^{-5}$ quoted by HFAG~\cite{HFAG}. The precision of this prediction is
limited by that of the input measurements and LQCD inputs, and can be
systematically improved with new data from Belle~II or LHCb.  

\begin{table}[t!]
\begin{tabular}{l|ccc}
\hline\hline
\multicolumn{1}{c|}{Scenario}		& $R(D)$  &  $R(D^*)$  &  Correlation \\ \hline
    $\lqcdGF$ 				& $0.292 \pm 0.005$ 	& $0.255 \pm 0.005$ 	& 41\% \\
    $\lqcdGFQ$ 	& $0.291 \pm 0.005$ 	& $0.255 \pm 0.003$ 	& 57\% \\
    $\normDDs$ 				& $0.273 \pm 0.016$ 	& $0.250 \pm 0.006$ 	& 49\% \\ 
    $\normDDsQ$ 	& $0.295 \pm 0.007$ 	& $0.255 \pm 0.004$ 	& 43\%\\
    $\lqcdXw$ 	& $0.298 \pm 0.003$ 	& $0.261 \pm 0.004$ 	& 19\% \\
    $\lqcdXwQ$ 	& $\bm{0.299 \pm 0.003}$ 	& $\bm{0.257 \pm 0.003}$ 	& $\bm{44}$\% \\
    $\lqcdXwQth$ & $0.306 \pm 0.005$ 	& $0.256 \pm 0.004$ 	& 33\%\\    
\hline
Data~\cite{HFAG}  &  $0.403 \pm 0.047$ & $0.310 \pm 0.017$ & $-23\%$ \\
\hline
Refs.~\cite{Aoki:2016frl,Lattice:2015rga,Na:2015kha} & $0.300 \pm 0.008$  & --- & --- \\
  Ref.~\cite{Bigi:2016mdz} 	& $0.299 \pm 0.003$ 	& ---  & --- \\
  Ref.~\cite{Fajfer:2012vx} 	&---  & $0.252 \pm 0.003$ 	& --- \\  
\hline\hline
\end{tabular}
\caption{The $R(D)$ and $R(D^*)$ predictions for our fit scenarios, the world
average of the data, and other theory predictions. The fit scenarios are
described in the text and in Table~\ref{tab:FitKey}. The bold numbers are our most precise predictions.}
\label{tab:RDRDs}
\end{table}

To derive a SM prediction for $R(D^*)$, Ref.~\cite{Fajfer:2012vx} used the
measured $R_2(1)$ form factor ratio~\cite{HFAG} and the QCDSR
predictions to obtain $R_0(1) =
1.14 \pm 0.11$. In comparison, our ``$\lqcdXwQ$" fit results yield
\begin{equation}
 R_0(1) = 1.17 \pm 0.02 \, ,  \qquad R_3(1) = 1.19 \pm 0.03 \,  .
\end{equation}
The precision on $R_0(1)$ improves five-fold compared to Ref.~\cite{Fajfer:2012vx} 
and is in good agreement. 

\begin{figure}[t!]
\centerline{\includegraphics[width=0.375\textwidth]{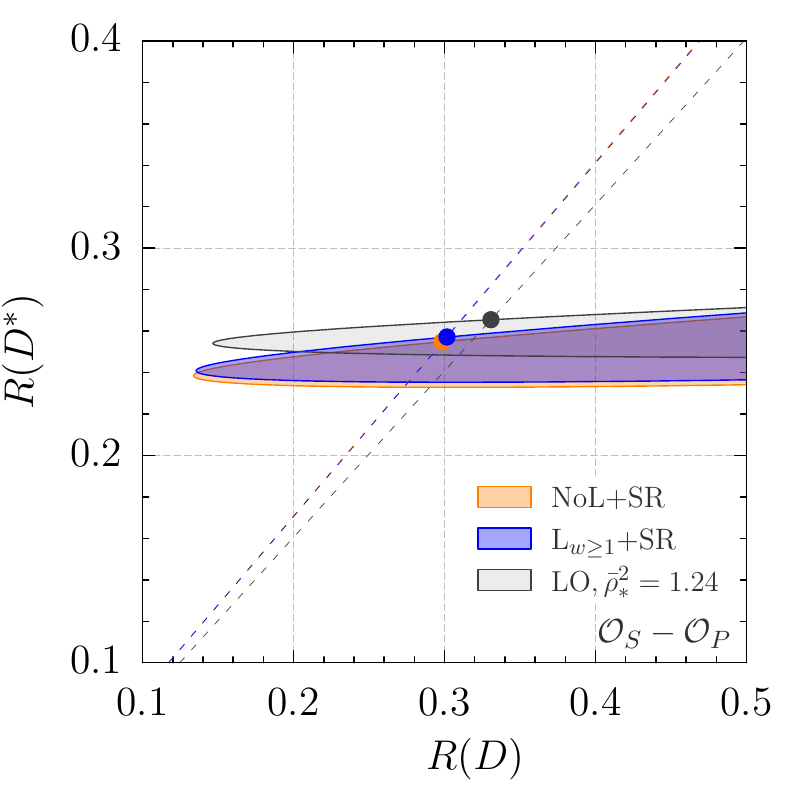} \hfil
  \includegraphics[width=0.375\textwidth]{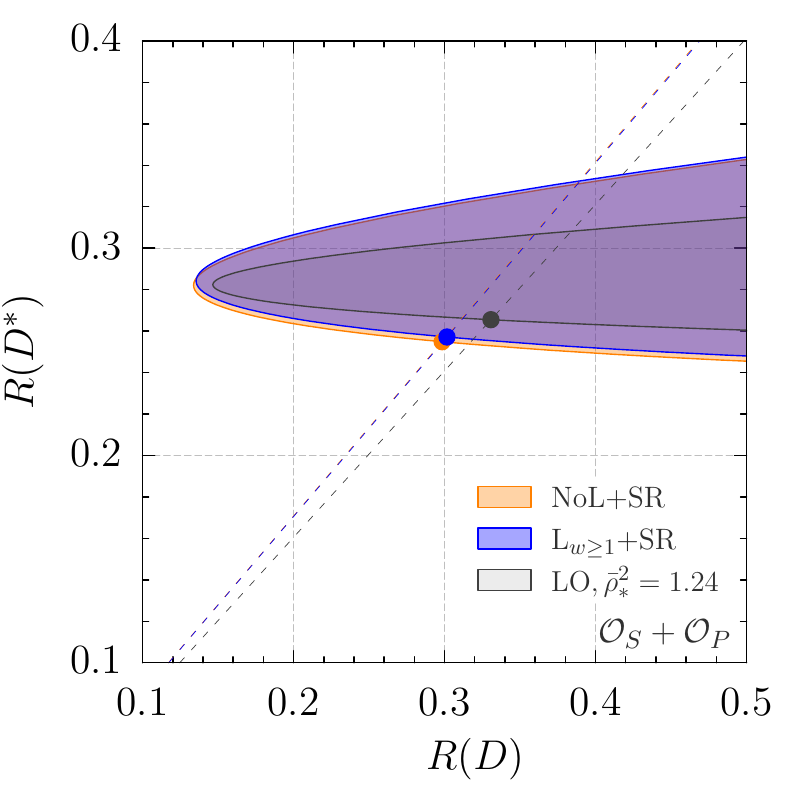}}
\centerline{\includegraphics[width=0.375\textwidth]{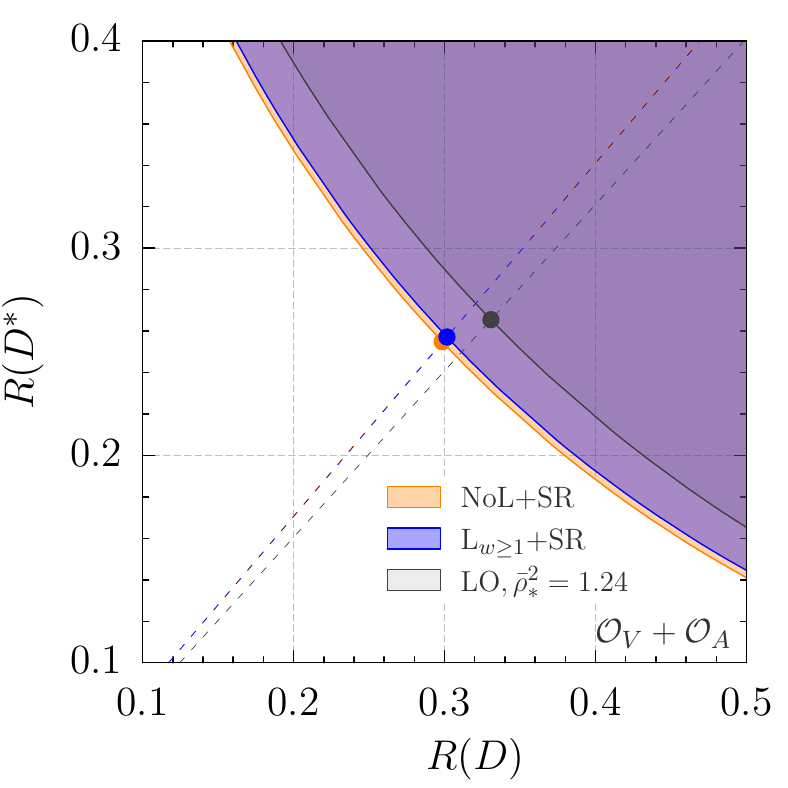} \hfil
  \includegraphics[width=0.375\textwidth]{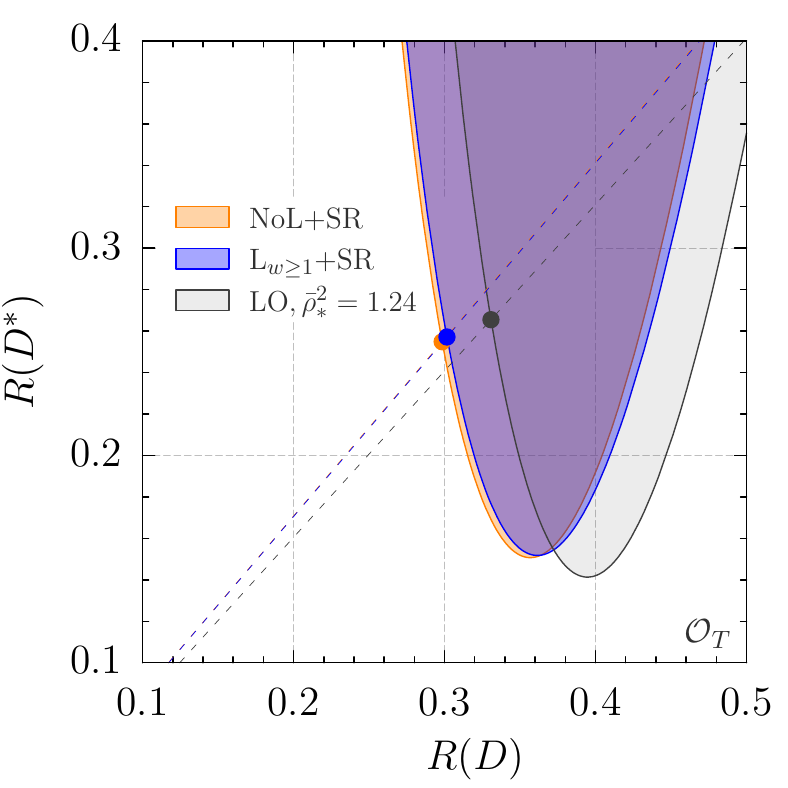}}
\vspace*{-10pt}
\caption{The allowed ranges of $R(D) - R(D^*)$, due to one of the new physics
operators in addition to the SM: $O_S - O_P$ (top left), $O_S + O_P$ (top right), 
$O_V + O_A$ (bottom left), $O_T$ (bottom right).}
\label{fig:NP}
\end{figure}

In Fig.~\ref{fig:NP} we illustrate the impacts NP might have on the allowed
$R(D) - R(D^*)$ regions, assuming the dominance of one new physics operator in a
standard four-Fermi basis. NP couplings are permitted to have an arbitrary
phase, generating allowed regions rather than single contours. We display the
allowed regions generated for the ``$\normDDsQ$'' best fit values; the ``$\lqcdXwQ$'' best fit values; and for leading order
contributions only, i.e., $\aS$, $\varepsilon_{c,b} \to 0$, with $\brhosq =
1.24$. The small variation between the ``$\normDDsQ$'' and ``$\lqcdXwQ$'' regions illustrates the good consistency 
of the predictions obtained with and without LQCD.
On each plot, we also include for comparison the corresponding contours (dashed lines) produced by a NP $O_V-O_A$ coupling.
The latter rescales $R(D)$ and $R(D^*)$ keeping their ratio fixed. Solid dots indicate the SM point for each case. For
scalar currents, if NP only contributes to $O_S$ ($O_P$) then only $R(D)$
($R(D^*)$) is affected in accordance with Eq.~(\ref{eqn:DPA})
(Eq.~(\ref{eqn:DsS})), respectively. We plot the allowed regions
for the $O_S\pm O_P$ linear combinations, which are also motivated by specific NP models.

\section{Summary and Outlook}
\label{sec:conc}

We performed a novel combined fit of the $\bdln$ and $\bdsln$ differential rates
and angular distributions, consistently including the HQET relations to
$\mathcal{O}(\lqcd/{m_{c,b}},\, \aS)$. Under various fit scenarios, that use or
omit lattice QCD and QCD sum rule predictions, we constrain the leading and
subleading Isgur-Wise functions.  We thus obtain strong constraints on all form
factors, and predictions for the form factor ratios $R_{1,2}$ as well as
$R(\dds)$, both in the SM and in arbitrary NP scenarios, valid at
$\mathcal{O}(\aS)$ and $\mathcal{O}(\lqcd/m_{c,b})$.  Our most precise
prediction for $R(\dds)$, in the ``$\lqcdXwQ$" fit, using the experimental data
and all lattice QCD and QCDSR inputs is
\begin{equation}
R(D) = 0.299 \pm 0.003\, , \qquad R(D^*) = 0.257 \pm 0.003 \, ,
\end{equation}
with a correlation of 44\%.  The same fit also yields
$|V_{cb}| = ( 39.3 \pm 1.0 ) \times 10^{-3}$,
which is in good agreement with existing exclusive determinations.
All possible $b \to c$ current form factors are derived at $\mathcal{O}( \lqcd/m_{c,b})$ 
and $\mathcal{O}(\aS)$, including those for a tensor current,
previously unavailable in the literature at this order. 
A lattice QCD calculation of the subleading Isgur-Wise functions, or even just
those which arise from the chromomagnetic term in the subleading HQET
Lagrangian ($\chi_{2,3}$), would be important to reduce hadronic uncertainties
in both SM and NP predictions, complementary to a long-awaited lattice
calculation of $R(D^*)$.

At the current level of experimental precision, our predictions agree up to mild
tensions with previous results, which neglected the HQET relations for the
uncertainties of the $\mathcal{O}(\lqcd/m_{c,b})$ terms.  Our fit results are
consistent with one another, and at the current level of precision we find no
inconsistencies between the data, lattice QCD results, and QCD sum rule
predictions.  Our fit using all available lattice QCD and QCD sum rule inputs
and HQET to order $\mathcal{O}(\aS, \lqcd/m_{c,b})$ yields the most precise
combined prediction for $R(D)$ and $R(D^*)$ to date.  However, in principle, our
fit need not require either lattice or sum rule input, and its precision can be
improved simply as the statistics of future data increases. 

The (moderate) tension between the measurements of $|V_{cb}|$ from inclusive and
exclusive semileptonic decays probably cannot be resolved with current data.
Understanding how the inclusive rate is made up from a sum of exclusive channels
has been unclear from the data for a long time~\cite{Richman:1995wm}, and
puzzles remain even in light of BaBar and Belle
measurements~\cite{Bernlochner:2012bc, Bernlochner:2014dca}.  A more detailed
examination of the effects of the unitarity constraints and the precision
extraction of $|V_{cb}|$ is the subject of ongoing work~\cite{Vcb:2017}.  We
are also implementing the full angular distributions of the measurable
particles~\cite{Ligeti:2016npd, Alonso:2016gym} into a software package, \texttt{hammer}~\cite{hammer_ichep, hammer_paper},
based on the state-of-the-art HQET predictions for all six $B\to D,
D^*,D^{**}$ decay modes.

\acknowledgments

We thank Marat Freytsis, Ben Grinstein and Aneesh Manohar for helpful
conversations. 
We thank Martin Jung for pointing out typos in Eqs.~(\ref{A1a}) and (\ref{A1b})
in an earlier version.
FB was supported by the DFG Emmy-Noether Grant No.\ BE 6075/1-1. 
FB thanks Kim Scott and Robert Michaud for the kind hospitality in
Houston where part of this work was carried out and inspiring conversations over
good wine and food. ZL and MP were supported in part by the  U.S.\ Department of
Energy under contract DE-AC02-05CH11231. DR acknowledges support from the
University of Cincinnati.

\appendix

\section{The \texorpdfstring{$\mathcal{O}(\aS)$}{Oas} corrections}
\label{app:aSC}

In this appendix we summarize the explicit expressions for the $C_{\Gamma}(w)$
functions defined in Eq.~\eqref{eqn:ascurrent}, calculated in
Ref.~\cite{Neubert:1992qq}. The following results use the $\overline{\rm MS}$
scheme and correspond to matching from QCD onto HQET at $\mu = \sqrt{m_c m_b}$,
\begin{subequations}\label{A1}
\begin{align}
C_{S} & = \frac{1}{3z(w - \wm)}\Big[2 z (w-\wm) \Omega(w) - (w-1) (z+1)^2
r(w) +  (z^2 -1) \ln z\Big], \label{A1a} \\
C_{P} & = \frac{1}{3z(w - \wm)}\Big[2 z (w-\wm) \Omega(w) - (w+1) (z-1)^2
r(w) + (z^2 -1) \ln z\Big], \label{A1b} \\
C_{V_1} & =  \frac{1}{6 z(w - \wm)}\Big[2(w+1)\big((3w-1) z - z^2 - 1\big) r(w) \nn \\
  & \qquad \qquad  \qquad \qquad + \big(12 z (\wm - w) - (z^2 - 1) \ln z\big) + 4 z (w - \wm) \Omega(w)\Big], \\
C_{V_2} & = \frac{-1}{6 z^2(w - \wm)^2}\Big[\big((4 w^2+2 w) z^2-(2 w^2+5 w-1) z - (w+1) z^3+2\big) r(w) \nn \\
  & \qquad \qquad \qquad  \qquad + z\big(2 (z-1) (\wm-w)+\big(z^2 - (4w -2) z+(3-2 w)\big) \ln z\big) \Big], \\
C_{V_3} & = \frac{1}{6z(w - \wm)^2}\Big[ \big((2 w^2+5w-1) z^2- (4 w^2 + 2w) z-2 z^3+w+1\big) r(w) \nn \\
  & \qquad \qquad \qquad \qquad + \big(2 z(z-1) (\wm-w)+((3-2 w) z^2+(2-4 w) z+1) \ln z \big)\Big], \\
C_{A_1} & = \frac{1}{6 z(w - \wm)}\Big[2 (w-1) \big((3 w+1) z- z^2 -1\big) r(w) \nn \\
  & \qquad \qquad \qquad \qquad + \big(12 z (\wm-w)-(z^2-1)\ln z\big)+4 z (w-\wm) \Omega(w)\Big], \\
C_{A_2} & = \frac{-1}{6 z^2(w - \wm)^2}\Big[  \big((4 w^2-2 w) z^2+(2 w^2-5 w-1) z+(1-w) z^3+2\big) r(w) \nn \\
  & \qquad \qquad \qquad \qquad +z \big(2 (z+1) (\wm-w)+\big(z^2 - (4 w+2) z+(2 w+3)\big) \ln z\big) \Big], \\
C_{A_3} & = \frac{1}{6 z(w - \wm)^2}\Big[\big(2 z^3 + (2 w^2-5 w-1) z^2+(4 w^2-2 w) z-w+1\big) r(w) \nn \\
  & \qquad \qquad \qquad \qquad +  \! \big(2 z (z+1) (\wm-w) - \big((2 w+3) z^2- (4 w+2) z+ 1\big) \ln z \big)\Big], \\	
C_{T_1} & = \frac{1}{3 z(w - \wm)}\Big[(w-1) \big((4 w+2) z-z^2-1\big) r(w) \nn\\*
  & \qquad \qquad \qquad \qquad  + \big(6 z (\wm- w) - (z^2-1)\ln z\big)+2 z (w-\wm) \Omega(w) \Big], \\ 
C_{T_2} &= \frac{2}{3 z(w - \wm)}\Big[(1-w z)r(w) +  z \ln z \Big], \\
C_{T_3} &= \frac{2}{3 (w - \wm)}\Big[(w-z)r(w) +  \ln z \Big],
\end{align}
\end{subequations}
and $C_{T_4} = 0$.  Here $z = m_c/m_b$, and the functions
\begin{multline}
	\Omega(w) \equiv \frac{w}{2\sqrt{w^2-1}}\big[ 2 \text{Li}_2(1 - w_- z) - 2\text{Li}_2(1-w_+ z) + \text{Li}_2(1 - w_+^2) - \text{Li}_2(1 - w_-^2)\big] \\ - w\, r(w) \ln z + 1\,,
\end{multline}
where $\text{Li}_2(x) = \int_x^0 \ln(1-t)/t\, \d t$ is the dilogarithm, and
\begin{equation}
	r(w) \equiv \frac{\ln w_+}{\sqrt{w^2-1}}\,, \qquad
	w_\pm \equiv w \pm \sqrt{w^2-1}\,, \qquad 
	\wm \equiv \frac12\, \big(z + 1/z \big)\,.
\end{equation}
At the zero recoil point, $w=1$, 
\begin{align}
\label{alphaexp0}
C_S(1) & = - \frac{2}{3} \,, \qquad  C_P(1) = \frac{2}{3} \,, \nn\\*
C_{V_1}(1)  & = - \frac{4}{3} - \frac{1+z}{1-z} \ln z , \quad
  C_{V_2}(1) = - \frac{2\,(1-z+z \ln z) }{ 3 (1-z)^2} \,, \quad
  C_{V_3}(1) = \frac{2z (1-z+\ln z) }{ 3 (1-z)^2} \,, \nn\\
C_{A_1}(1)  & = - \frac{8}{3} - \frac{1+z}{1-z}\, \ln z \,,\nn \\
C_{A_2}(1) & = - \frac{2\,[3-2z-z^2+(5-z)z\ln z] }{ 3(1-z)^3} \,, \qquad  C_{A_3}(1) = \frac{2z\, [1+2z-3z^2+(5z-1)\ln z] }{ 3 (1-z)^3} \,, \nn\\
C_{T_1}(1) & = - \frac{8}{3} - \frac{4(1+z)}{ 3(1-z)}\, \ln z \,, \qquad  C_{T_2}(1) = 2\, C_{V_2}(1)\,, \qquad C_{T_3}(1) = -2\, C_{V_3}(1) \,.
\end{align}
Finally, for arbitrary matching scale $\mu$, one should add to Eqs.~(\ref{A1})
the terms
\begin{subequations}
\begin{align}
C_{S,P}^{(\mu^2)} & = C_{S,P}^{(m_bm_c)} 
  - \frac13\, [2 w\, r(w) +1] \ln(m_cm_b/\mu^2)\,, \\
C_{V_1,A_1}^{(\mu^2)} & = C_{V_1,A_1}^{(m_bm_c)}
  - \frac23\, [w\, r(w) - 1] \ln(m_cm_b/\mu^2)\,, \\
C_{T_1}^{(\mu^2)} & = C_{T_1}^{(m_bm_c)}
  - \frac13\, [2 w\, r(w) -3] \ln(m_cm_b/\mu^2)\,,
\end{align}
\end{subequations}
and all other $C_{\Gamma_j}^{(\mu^2)} = C_{\Gamma_j}^{(m_bm_c)}$, for $j\ge 2$.

\section{Dull Correlations}
\label{app:DC}
\newcommand{\scaletab}[1]{\scalebox{0.7}{\parbox{\linewidth}{#1}}}
The correlation matrices for the fit scenarios are given in Tables~\ref{tab:corrfirst}-\ref{tab:corrlast}.
\begin{table}[ht!] \scaletab{\newcolumntype{C}{ >{\centering\arraybackslash $} c <{$}}
\begin{tabular}{C|CCCCCCCCCC}
 \hline\hline
& |V_{cb}| & \mathcal{G}(1) & \mathcal{F}(1) & \brhosq & \hat \chi_{2}(1) & \hat \chi_{2}'(1) & \hat \chi_{3}'(1) & \eta(1) & m_{b}^{1S} & \delta m_{bc}  \\ \hline
|V_{cb}|& 1.00 & -0.16 & -0.18 & 0.30 & -0.13 & 0.28 & 0.11 & 0.04  & -0.01 & 0.00  \\
\mathcal{G}(1)& -0.16 & 1.00 & 0.06 & -0.11 & 0.03 & -0.04 & -0.09 & -0.23 & 0.00 & -0.00  \\
\mathcal{F}(1)& -0.18 & 0.06 & 1.00 & 0.18 & -0.00 & 0.08 & 0.21 & -0.02 &  0.01 & -0.00  \\
\brhosq& 0.30 & -0.11 & 0.18 & 1.00 & 0.67 & -0.47 & 0.82 & 0.13 & -0.16 & 0.01  \\
\hat \chi_{2}(1)& -0.13 & 0.03 & -0.00 & 0.67 & 1.00 & -0.87 & 0.82 & -0.11  & 0.07 & -0.01  \\
\hat \chi_{2}'(1)& 0.28 & -0.04 & 0.08 & -0.47 & -0.87 & 1.00 & -0.47 & 0.01  & 0.01 & -0.00  \\
\hat \chi_{3}'(1)& 0.11 & -0.09 & 0.21 & 0.82 & 0.82 & -0.47 & 1.00 & -0.12  & 0.12 & -0.02  \\
\eta(1)& 0.04 & -0.23 & -0.02 & 0.13 & -0.11 & 0.01 & -0.12 & 1.00  & -0.52 & 0.05  \\
m_{b}^{1S}& -0.01 & 0.00 & 0.01 & -0.16 & 0.07 & 0.01 & 0.12 & -0.52  & 1.00 & 0.00  \\
\delta m_{bc}& 0.00 & -0.00 & -0.00 & 0.01 & -0.01 & -0.00 & -0.02 & 0.05  & 0.00 & 1.00  \\
\hline\hline
\end{tabular}
} \caption{The correlations of the ``$\lqcdGF$'' fit scenario.} \label{tab:corrfirst}   \end{table} \vspace*{30pt}
\begin{table}[ht!] \scaletab{\newcolumntype{C}{ >{\centering\arraybackslash $} c <{$}}
\begin{tabular}{C|CCCCCCCCCCC}
 \hline\hline
& |V_{cb}| & \mathcal{G}(1) & \mathcal{F}(1) & \brhosq & \hat \chi_{2}(1) & \hat \chi_{2}'(1) & \hat \chi_{3}'(1) & \eta(1) & \eta'(1) & m_{b}^{1S} & \delta m_{bc}  \\ \hline
|V_{cb}|& 1.00 & -0.12 & -0.32 & 0.48 & -0.02 & 0.02 & 0.14 & 0.05 & 0.02 & -0.02 & 0.00  \\
\mathcal{G}(1)& -0.12 & 1.00 & 0.14 & -0.05 & 0.04 & 0.01 & -0.14 & -0.23 & 0.09 & -0.00 & -0.00  \\
\mathcal{F}(1)& -0.32 & 0.14 & 1.00 & 0.04 & -0.07 & -0.01 & 0.24 & -0.02 & -0.11 & -0.03 & 0.01  \\
\brhosq& 0.48 & -0.05 & 0.04 & 1.00 & -0.09 & -0.04 & 0.57 & 0.32 & 0.08 & -0.45 & 0.04  \\
\hat \chi_{2}(1)& -0.02 & 0.04 & -0.07 & -0.09 & 1.00 & -0.03 & 0.17 & -0.06 & -0.20 & 0.04 & -0.00  \\
\hat \chi_{2}'(1)& 0.02 & 0.01 & -0.01 & -0.04 & -0.03 & 1.00 & 0.06 & -0.02 & -0.09 & 0.01 & -0.00  \\
\hat \chi_{3}'(1)& 0.14 & -0.14 & 0.24 & 0.57 & 0.17 & 0.06 & 1.00 & 0.08 & 0.38 & -0.03 & 0.00  \\
\eta(1)& 0.05 & -0.23 & -0.02 & 0.32 & -0.06 & -0.02 & 0.08 & 1.00 & -0.14 & -0.48 & 0.05  \\
\eta'(1)& 0.02 & 0.09 & -0.11 & 0.08 & -0.20 & -0.09 & 0.38 & -0.14 & 1.00 & 0.08 & -0.01  \\
m_{b}^{1S}& -0.02 & -0.00 & -0.03 & -0.45 & 0.04 & 0.01 & -0.03 & -0.48 & 0.08 & 1.00 & 0.01  \\
\delta m_{bc}& 0.00 & -0.00 & 0.01 & 0.04 & -0.00 & -0.00 & 0.00 & 0.05 & -0.01 & 0.01 & 1.00  \\
\hline\hline
\end{tabular}
}  \caption{The correlations of the ``$\lqcdGFQ$'' fit scenario.}  \end{table} \vspace*{30pt}
\begin{table}[ht!] \scaletab{\newcolumntype{C}{ >{\centering\arraybackslash $} c <{$}}
\begin{tabular}{C|CCCCCCCC}
 \hline\hline
& \brhosq & \hat \chi_{2}(1) & \hat \chi_{2}'(1) & \hat \chi_{3}'(1) & \eta(1) & m_{b}^{1S} & \delta m_{bc}  \\ \hline
\brhosq& 1.00 & -0.22 & -0.18 & -0.03 & 0.46 & -0.22 & 0.01  \\
\hat \chi_{2}(1)& -0.22 & 1.00 & -0.41 & 0.94 & -0.92 & 0.33 & -0.03  \\
\hat \chi_{2}'(1) & -0.18 & -0.41 & 1.00 & -0.19 & 0.08 & -0.02 & -0.00  \\
\hat \chi_{3}'(1) & -0.03 & 0.94 & -0.19 & 1.00 & -0.88 & 0.32 & -0.03  \\
\eta(1) & 0.46 & -0.92 & 0.08 & -0.88 & 1.00  & -0.35 & 0.02  \\
m_{b}^{1S} & -0.22 & 0.33 & -0.02 & 0.32 & -0.35  & 1.00 & 0.00  \\
\delta m_{bc}& 0.01 & -0.03 & -0.00 & -0.03 & 0.02 & 0.00 & 1.00  \\
\hline\hline
\end{tabular}
}  \caption{The correlations of the ``$\normDDs$'' fit scenario.} \end{table} \vspace*{30pt}
\begin{table}[ht!] \scaletab{\newcolumntype{C}{ >{\centering\arraybackslash $} c <{$}}
\begin{tabular}{C|CCCCCCCC}
 \hline\hline
& \brhosq & \hat \chi_{2}(1) & \hat \chi_{2}'(1) & \hat \chi_{3}'(1) & \eta(1) & \eta'(1) & m_{b}^{1S} & \delta m_{bc}  \\ \hline
\brhosq & 1.00 & -0.15 & -0.07 & 0.57 & 0.44 & -0.11 & -0.31 & 0.03  \\
\hat \chi_{2}(1) & -0.15 & 1.00 & -0.02 & 0.07 & -0.15 & -0.09 & 0.02 & -0.00  \\
\hat \chi_{2}'(1) & -0.07 & -0.02 & 1.00 & 0.03 & -0.07 & -0.05 & 0.00 & -0.00  \\
\hat \chi_{3}'(1) & 0.57 & 0.07 & 0.03 & 1.00 & 0.17 & 0.16 & 0.00 & -0.00  \\
\eta(1) & 0.44 & -0.15 & -0.07 & 0.17 & 1.00 & -0.40 & 0.09 & -0.01  \\
\eta'(1)& -0.11 & -0.09 & -0.05 & 0.16 & -0.40 & 1.00 & 0.02 & -0.00  \\
m_{b}^{1S} & -0.31 & 0.02 & 0.00 & 0.00 & 0.09 & 0.02 & 1.00 & 0.01  \\
\delta m_{bc} & 0.03 & -0.00 & -0.00 & -0.00 & -0.01 & -0.00 & 0.01 & 1.00  \\
\hline\hline
\end{tabular}
} \caption{The correlations of the ``$\normDDsQ$'' fit scenario.} \end{table} \vspace*{30pt}
\begin{table}[ht!] \scaletab{\newcolumntype{C}{ >{\centering\arraybackslash $} c <{$}}
\begin{tabular}{C|CCCCCCCCCCC}
 \hline\hline
& |V_{cb}| \times 10^{3} & \mathcal{G}(1) & \mathcal{F}(1) & \brhosq & \chi_{2}(1) & \chi_{2}' & \chi_{3}' & \eta(1)& m_{b}^{1S} & \delta m_{bc}  \\ \hline
|V_{cb}| \times 10^{3}& 1.00 & -0.30 & -0.16 & 0.18 & -0.13 & 0.28 & 0.07 & 0.01 & 0.00 & 0.00  \\
\mathcal{G}(1)& -0.30 & 1.00 & 0.08 & -0.28 & -0.04 & -0.04 & -0.16 & -0.23  & 0.01 & -0.00  \\
\mathcal{F}(1)& -0.16 & 0.08 & 1.00 & 0.38 & 0.18 & -0.10 & 0.32 & 0.00  & 0.01 & -0.00  \\
\brhosq& 0.18 & -0.28 & 0.38 & 1.00 & 0.64 & -0.44 & 0.80 & 0.18  & -0.22 & 0.01  \\
\hat \chi_{2}(1)& -0.13 & -0.04 & 0.18 & 0.64 & 1.00 & -0.79 & 0.89 & -0.17  & 0.21 & -0.03  \\
\hat \chi_{2}'(1)& 0.28 & -0.04 & -0.10 & -0.44 & -0.79 & 1.00 & -0.48 & 0.05  & -0.13 & 0.02  \\
\hat \chi_{3}'(1)& 0.07 & -0.16 & 0.32 & 0.80 & 0.89 & -0.48 & 1.00 & -0.12 & 0.18 & -0.03  \\
\eta(1)& 0.01 & -0.23 & 0.00 & 0.18 & -0.17 & 0.05 & -0.12 & 1.00 & -0.54 & 0.05  \\
m_{b}^{1S}& 0.00 & 0.01 & 0.01 & -0.22 & 0.21 & -0.13 & 0.18 & -0.54 & 1.00 & 0.01  \\
\delta m_{bc}& 0.00 & -0.00 & -0.00 & 0.01 & -0.03 & 0.02 & -0.03 & 0.05 & 0.01 & 1.00  \\
\hline\hline
\end{tabular}
} \caption{The correlations of the ``$\lqcdXw$'' fit scenario.} \end{table} \vspace*{30pt}
\begin{table}[ht!] \scaletab{\newcolumntype{C}{ >{\centering\arraybackslash $} c <{$}}
\begin{tabular}{C|CCCCCCCCCCC}
 \hline\hline
& |V_{cb}| & \mathcal{G}(1) & \mathcal{F}(1) & \brhosq & \hat \chi_{2}(1) & \hat \chi_{2}'(1) & \hat \chi_{3}'(1) & \eta(1) & \eta'(1) & m_{b}^{1S} & \delta m_{bc}  \\ \hline
|V_{cb}|& 1.00 & -0.27 & -0.28 & 0.33 & -0.05 & 0.00 & 0.21 & 0.03 & 0.13 & -0.01 & 0.00  \\
\mathcal{G}(1)& -0.27 & 1.00 & 0.24 & -0.26 & 0.07 & 0.02 & -0.22 & -0.22 & -0.27 & -0.02 & 0.00  \\
\mathcal{F}(1)& -0.28 & 0.24 & 1.00 & 0.15 & -0.06 & -0.02 & 0.25 & -0.03 & -0.19 & -0.01 & 0.00  \\
\brhosq& 0.33 & -0.26 & 0.15 & 1.00 & -0.13 & -0.08 & 0.72 & 0.35 & 0.13 & -0.49 & 0.05  \\
\hat \chi_{2}(1)& -0.05 & 0.07 & -0.06 & -0.13 & 1.00 & -0.04 & 0.25 & -0.10 & -0.11 & 0.06 & -0.01  \\
\hat \chi_{2}'(1)& 0.00 & 0.02 & -0.02 & -0.08 & -0.04 & 1.00 & 0.07 & -0.05 & -0.06 & 0.04 & -0.00  \\
\hat \chi_{3}'(1)& 0.21 & -0.22 & 0.25 & 0.72 & 0.25 & 0.07 & 1.00 & 0.16 & 0.19 & -0.06 & 0.01  \\
\eta(1)& 0.03 & -0.22 & -0.03 & 0.35 & -0.10 & -0.05 & 0.16 & 1.00 & 0.05 & -0.48 & 0.05  \\
\eta'(1)& 0.13 & -0.27 & -0.19 & 0.13 & -0.11 & -0.06 & 0.19 & 0.05 & 1.00 & 0.04 & 0.00  \\
m_{b}^{1S}& -0.01 & -0.02 & -0.01 & -0.49 & 0.06 & 0.04 & -0.06 & -0.48 & 0.04 & 1.00 & 0.01  \\
\delta m_{bc}& 0.00 & 0.00 & 0.00 & 0.05 & -0.01 & -0.00 & 0.01 & 0.05 & 0.00 & 0.01 & 1.00  \\
\hline\hline
\end{tabular}
} \caption{The correlations of the ``$\lqcdXwQ$'' fit scenario.} \end{table} \vspace*{30pt}
\begin{table}[ht!] \scaletab{\newcolumntype{C}{ >{\centering\arraybackslash $} c <{$}}
\begin{tabular}{C|CCCCCCCCCC}
 \hline\hline
 & \mathcal{G}(1) & \mathcal{F}(1) & \brhosq & \hat \chi_{2}(1) & \hat \chi_{2}'(1) & \hat \chi_{3}'(1) & \eta(1) & \eta'(1) & m_{b}^{1S} & \delta m_{bc}  \\ \hline
\mathcal{G}(1)& 1.00 & 0.00 & -0.15 & 0.01 & -0.00 & -0.02 & -0.25 & -0.40 & 0.01 & -0.00  \\
\mathcal{F}(1) & 0.00 & 1.00 & -0.00 & -0.00 & -0.00 & -0.00 & -0.00 & -0.00 & -0.00 & -0.00  \\
\brhosq & -0.15 & -0.00 & 1.00 & -0.27 & -0.13 & 0.81 & 0.08 & -0.07 & -0.24 & 0.02  \\
\hat \chi_{2}(1) & 0.01 & -0.00 & -0.27 & 1.00 & 0.00 & 0.01 & 0.01 & 0.03 & -0.01 & 0.00  \\
\hat \chi_{2}'(1)& -0.00 & -0.00 & -0.13 & 0.00 & 1.00 & -0.01 & -0.01 & 0.01 & 0.01 & -0.00  \\
\hat \chi_{3}'(1) & -0.02 & -0.00 & 0.81 & 0.01 & -0.01 & 1.00 & -0.02 & -0.09 & 0.04 & -0.00  \\
\eta(1) & -0.25 & -0.00 & 0.08 & 0.01 & -0.01 & -0.02 & 1.00 & 0.11 & -0.48 & 0.04  \\
\eta'(1) & -0.40 & -0.00 & -0.07 & 0.03 & 0.01 & -0.09 & 0.11 & 1.00 & 0.07 & -0.01  \\
m_{b}^{1S} & 0.01 & -0.00 & -0.24 & -0.01 & 0.01 & 0.04 & -0.48 & 0.07 & 1.00 & 0.00  \\
\delta m_{bc} & -0.00 & -0.00 & 0.02 & 0.00 & -0.00 & -0.00 & 0.04 & -0.01 & 0.00 & 1.00  \\
\hline\hline
\end{tabular}
} \caption{The correlations of the ``$\lqcdXwQth$'' fit scenario.} \label{tab:corrlast}  \end{table}

\FloatBarrier
\bibliographystyle{apsrev4-1}

\end{document}